\documentclass[prd,superscriptaddress,nofootinbib]{revtex4}
\usepackage{amsmath,amssymb,epsfig}
\usepackage[hypertex]{hyperref}
\usepackage{natbib,ifthen}
\usepackage{bm}
\usepackage{xcolor}

\begin{document}

\newcommand{\fixme}[1]{{\textbf{Fixme: #1}}}
\newcommand{\detD}{{\det\!\cld}}
\newcommand{\clh}{\mathcal{H}}
\newcommand{\ud}{{\rm d}}
\renewcommand{\eprint}[1]{\href{http://arxiv.org/abs/#1}{#1}}
\newcommand{\adsurl}[1]{\href{#1}{ADS}}
\newcommand{\ISBN}[1]{\href{http://cosmologist.info/ISBN/#1}{ISBN: #1}}
\newcommand{\jcap}{J.\ Cosmol.\ Astropart.\ Phys.}
\newcommand{\mnras}{Mon.\ Not.\ R.\ Astron.\ Soc.}
\newcommand{\progress}{Rep.\ Prog.\ Phys.}
\newcommand{\prlett}{Phys.\ Rev.\ Lett.}
\newcommand{\procspie}{Proc.\ SPIE}
\newcommand{\na}{New Astronomy}
\newcommand{\apjl}{ApJ.\ Lett.}
\newcommand{\physrep}{Physics Reports}
\newcommand{\aap}{A\&A}
\newcommand{\aapr}{A\&A Rev.}
\newcommand{\vort}{\varpi}
\newcommand\ba{\begin{eqnarray}}
\newcommand\ea{\end{eqnarray}}
\newcommand\be{\begin{equation}}
\newcommand\ee{\end{equation}}
\newcommand\lagrange{{\cal L}}
\newcommand\cll{{\cal L}}
\newcommand\cln{{\cal N}}
\newcommand\clx{{\cal X}}
\newcommand\clz{{\cal Z}}
\newcommand\clv{{\cal V}}
\newcommand\cld{{\cal D}}
\newcommand\clt{{\cal T}}

\newcommand{\ThreeJSymbol}[6]{\begin{pmatrix}
#1 & #3 & #5 \\
#2 & #4 & #6
 \end{pmatrix}}

\newcommand\clo{{\cal O}}
\newcommand{\cla}{{\cal A}}
\newcommand{\clp}{{\cal P}}
\newcommand{\clr}{{\cal R}}
\newcommand{\uD}{{\mathrm{D}}}
\newcommand{\calE}{{\cal E}}
\newcommand{\calB}{{\cal B}}
\newcommand{\curl}{\,\mbox{curl}\,}
\newcommand\del{\nabla}
\newcommand\Tr{{\rm Tr}}
\newcommand\half{{\frac{1}{2}}}
\newcommand\fourth{{1\over 8}}
\newcommand\bibi{\bibitem}
\newcommand{\kf}{\beta}
\newcommand{\kfprod}{\alpha}
\newcommand\calS{{\cal S}}
\renewcommand\H{{\cal H}}
\newcommand\K{{\rm K}}
\newcommand\mK{{\rm mK}}
\newcommand\km{{\rm km}}
\newcommand\synch{\text{syn}}
\newcommand\opacity{\tau_c^{-1}}

\newcommand{\Psil}{\Psi_l}
\newcommand{\bsigma}{{\bar{\sigma}}}
\newcommand{\bI}{\bar{I}}
\newcommand{\bq}{\bar{q}}
\newcommand{\bv}{\bar{v}}
\renewcommand\P{{\cal P}}
\newcommand{\numfrac}[2]{{\textstyle \frac{#1}{#2}}}

\newcommand{\la}{\langle}
\newcommand{\ra}{\rangle}

\newcommand{\Omtot}{\Omega_{\mathrm{tot}}}
\newcommand\xx{\mbox{\boldmath $x$}}
\newcommand{\phpr} {\phi`}
\newcommand{\gam}{\gamma_{ij}}
\newcommand{\sqgam}{\sqrt{\gamma}}
\newcommand{\delk}{\Delta+3{\K}}
\newcommand{\dph}{\delta\phi}
\newcommand{\om} {\Omega}
\newcommand{\dom}{\delta^{(3)}\left(\Omega\right)}
\newcommand{\rar}{\rightarrow}
\newcommand{\Rar}{\Rightarrow}
\newcommand\gsim{ \lower .75ex \hbox{$\sim$} \llap{\raise .27ex \hbox{$>$}} }
\newcommand\lsim{ \lower .75ex \hbox{$\sim$} \llap{\raise .27ex \hbox{$<$}} }
\newcommand\bigdot[1] {\stackrel{\mbox{{\huge .}}}{#1}}
\newcommand\bigddot[1] {\stackrel{\mbox{{\huge ..}}}{#1}}
\newcommand{\Mpc}{\text{Mpc}}
\newcommand{\Al}{{A_l}}
\newcommand{\Bl}{{B_l}}
\newcommand{\eAl}{e^\Al}
\newcommand{\ix}{{(i)}}
\newcommand{\ixp}{{(i+1)}}
\renewcommand{\k}{\beta}
\newcommand{\HD}{\mathrm{D}}

\newcommand{\nonflat}[1]{#1}
\newcommand{\Cgl}{C_{\text{gl}}}
\newcommand{\Cgltwo}{C_{\text{gl},2}}
\newcommand{\He}{{\rm{He}}}
\newcommand{\Mhz}{{\rm MHz}}
\newcommand{\vx}{{\mathbf{x}}}
\newcommand{\ve}{{\mathbf{e}}}
\newcommand{\vv}{{\mathbf{v}}}
\newcommand{\vk}{{\mathbf{k}}}
\newcommand{\vn}{{\mathbf{n}}}

\newcommand{\vnhat}{{\hat{\mathbf{n}}}}
\newcommand{\vkhat}{{\hat{\mathbf{k}}}}
\newcommand{\taueps}{{\tau_\epsilon}}

\newcommand{\vgrad}{{\mathbf{\nabla}}}
\newcommand{\fbarln}{\bar{f}_{,\ln\epsilon}(\epsilon)}


\title[Cosmic velocities with cross-correlation dipoles]{Measuring cosmic velocities with 21\,cm intensity mapping and galaxy redshift survey cross-correlation dipoles}

\author{Alex Hall}
\email{ahall@roe.ac.uk}
\affiliation{Institute for Astronomy, University of Edinburgh, Royal Observatory, Blackford Hill, Edinburgh, EH9 3HJ, U.K.}

\author{Camille Bonvin}
\email{camille.bonvin@unige.ch}
\affiliation{D\'{e}partement de Physique Th\'{e}orique \& Center for Astroparticle Physics, Universit\'{e} de Gen\`{e}ve, 24 Quai E. Ansermet, 1211 Gen\`{e}ve 4, Switzerland.}



\begin{abstract}
We investigate the feasibility of measuring the effects of peculiar velocities in large-scale structure using the dipole of the redshift-space cross-correlation function. We combine number counts of galaxies with brightness-temperature fluctuations from 21\,cm intensity mapping, demonstrating that the dipole may be measured at modest significance ($\lesssim 2\sigma$) by combining the upcoming radio survey CHIME with the future redshift surveys of DESI and Euclid. More significant measurements ($\lesssim~10\sigma$) will be possible by combining intensity maps from the SKA with these of DESI or Euclid, and an even higher significance measurement ($\lesssim 100\sigma$) may be made by combining observables completely internally to the SKA. We account for effects such as contamination by wide-angle terms, interferometer noise and beams in the intensity maps, non-linear enhancements to the power spectrum, stacking multiple populations, sensitivity to the magnification slope, and the possibility that number counts and intensity maps probe the same tracers. We also derive a new expression for the covariance matrix of multi-tracer redshift-space correlation function estimators with arbitrary orientation weights, which may be useful for upcoming surveys aiming at measuring redshift-space clustering with multiple tracers.

\end{abstract}

\maketitle

\section{Introduction}
\label{sec:intro}

Large-scale peculiar velocities have long been a rich source of information on the physics of structure formation in the Universe~\citep[e.g.][]{Peebles, 1987MNRAS.227....1K, 1988lsmu.book.....R, 1999ApJ...515L...1F}. However, the direct measurement of peculiar velocity is rendered challenging by the necessity of obtaining an independent measure of the distance to a galaxy, such that the uniform Hubble flow may be subtracted from the measured redshift. This is usually achieved by using empirical relationships such as the Fundamental Plane to measure proxies for the intrinsic luminosity, and hence derive velocities via a luminosity distance~\citep[e.g.][]{2014MNRAS.445.2677S}. With these in hand, constraints may be placed on the standard cosmological model and its extensions~\citep[e.g.][]{2014MNRAS.444.3926J}. However, such measurements require high-resolution spectra in order to measure velocity dispersions, and are hence constrained to fairly low redshifts. 

An alternative approach to measuring peculiar velocities is provided by their impact on the clustering statistics of galaxies. In~\citep{1987MNRAS.227....1K,1989MNRAS.236..851L,1992ApJ...385L...5H} it was shown that velocities contribute to the observed clustering of matter via the transformation between real and redshift space. These redshift-space distortions, which modify the observed volume of the pixels in which the observer counts galaxies, can be comparable in magnitude to the real-space clustering caused by the large-scale dark matter density field, especially if accurate redshift information for the tracers is available~\citep{1998ASSL..231..185H}. This effect has been measured with high significance in two-point statistics of the galaxy overdensity field~\citep{2001Natur.410..169P,2014MNRAS.443.1065B}.

As well as the redshift-space distortion effect, there are signatures of peculiar velocities in the observed galaxy overdensity from the Doppler effect. The reason for this is that when peculiar velocities are present, a galaxy observed at a given redshift will have a different conformal distance from the observer to the FRW expectation corresponding to that redshift. A galaxy with positive peculiar velocity will actually be closer to us (in terms of conformal distance) than an FRW calculation would suggest. Since the background matter density is decaying with time due to expansion, this galaxy resides in a patch of spacetime having lower mean density than the sky-average. Since the peculiar velocity varies over the sky, this results in a contribution to the observed inhomogeneity in the matter distribution from peculiar velocities~\citep{1987MNRAS.227....1K,2008MNRAS.389..292P}. Moreover, since observations are made on our past light-cone, a wrong estimation of the conformal distance to the galaxy results in a wrong estimation of the comoving time at which the observed photons have been emitted. Since both the Hubble flow and the peculiar velocities evolve with time, this results in additional distortions to the observed galaxy overdensity~\citep{Yoo:2010ni,Yoo:2009au,2011PhRvD..84d3516C, 2011PhRvD..84f3505B}. Unfortunately, these beyond-standard Doppler terms are generally much smaller than the density and redshift-space distortion terms, making their detection very challenging.


Recently, a potential method for isolating the effects of peculiar velocities has been proposed~\citep{2014PhRvD..89h3535B}. This method uses the fact that peculiar velocities are among the few contributions to the observed number counts that source an \emph{antisymmetric} part to the two-point correlation function of matter tracers, i.e. odd with respect to a flip in sign of the pair-separation vector. Additional sources of an antisymmetric correlation function come from the effect of gravitational redshift~\citep{2011Natur.477..567W,2013MNRAS.434.3008C}, although these may subsumed into velocity terms with use of the Euler equation~\citep{2014CQGra..31w4002B}.

The antisymmetric correlation function thus represents a promising route to measuring large-scale motions in the galaxy distribution. As is evident from symmetry considerations, the effect is only non-zero when distinct populations of tracers having different biases or magnification slopes are cross-correlated. This technique is thus a `multi-tracer' method - such methods have been shown to be very promising avenues for beating down cosmic variance~\citep{2009JCAP...10..007M}. Cross-correlations have the further advantage of being unbiased to many of the systematics that affect the auto-correlation, such as foregrounds and Poisson noise.

We thus require two distinct populations to be isolated from a galaxy survey, such as red/blue galaxies or bright/faint galaxies. A measurement has already been attempted in~\citep{2015arXiv151203918G}. Alternatively, we may combine galaxy number counts with a distinct tracer from a different survey. In this work, we combine redshift surveys with another tracer of large-scale structure, the 21\,cm brightness temperature fluctuation measured with intensity mapping. This technique has the advantage of having very precisely measured redshifts covering large cosmic volumes~\citep{2009astro2010S.234P}. Additionally, at the low redshifts we consider in this work, 21\,cm emitters are expected to trace the linear matter density on large scales, with relatively small contributions from ionization-fraction perturbations and non-Gaussianity~\citep{2009MNRAS.397.1926W}.

We will investigate to what extent the cross-correlation dipole may be measured with combinations of galaxy number counts and 21\,cm intensity mapping from current and future surveys. It may be shown that antisymmetry in the cross-correlation function corresponds to \emph{imaginary} terms in the power spectrum~\citep{2009JCAP...11..026M,Yoo:2012se}. The relative merits of correlation functions over power spectra have been discussed at length in the literature~\citep[e.g.][]{1994ApJ...426...23F, 2014JCAP...06..008T} so we will not belabour the subject here, merely noting that the cross-power spectrum provides an alternative method for measuring the effects discussed in this paper.

The structure of this paper is as follows. In Section~\ref{sec:sig} we write down an expression for the signal and discuss potential contaminants. In Section~\ref{sec:var} we write down an estimator for this signal and derive an expression for its covariance matrix. In Section~\ref{sec:surveys} we describe the surveys we consider in our forecasts, which are presented in Section~\ref{sec:forecasts} and tabulated in Appendix~\ref{app:sn}. We conclude in Section~\ref{sec:concs}.

We will set the speed of light to unity throughout this work, and assume a fiducial flat $\Lambda$CDM cosmology given by $(\Omega_b h^2, \Omega_c h^2, h, 10^9 A_s,n_s) = (0.0223,0.1056,0.73,1.8,1.0)$.

\section{Cross-correlation dipoles}
\label{sec:sig}
In this section we derive an expression for the redshift-space cross-correlation dipole assuming linear theory, which should be valid on the scales of interest. Non-linear corrections are discussed in Secion~\ref{subsec:systematics}. The derivation closely follows that of~\citep{2014PhRvD..89h3535B} for galaxy number counts, and we extend the analysis of that work to include fluctuations in the 21\,cm brightness temperature as an additional observable.

The expression for the observed fluctuation in galaxy number counts has been derived in~\citep{Yoo:2010ni,Yoo:2009au, 2011PhRvD..84d3516C,2011PhRvD..84f3505B}. The dominant terms in the overdensity of counts in a pixel at redshift $z$ in direction $\hat{\mathbf{n}}$ having apparent magnitude $m$ less than the survey limit $m_*$ are
\begin{align}
\label{eq:numcount}
\Delta_N(\hat{\mathbf{n}},z,m<m_*) &\approx b_N(z) D(L > L_*) - \frac{1}{\mathcal{H}}\hat{\mathbf{n}} \cdot \frac{ \partial \mathbf{v}}{\partial r}
+[5s(m_*,z) - 2]\int_0^r \mathrm{d}r' \frac{r-r'}{2rr'}\Delta_\Omega (\Phi+\Psi) \\
&-\left[5s(m_*,z)+\frac{2-5s(m_*,z)}{r\mathcal{H}}+\frac{\dot{\mathcal{H}}}{\mathcal{H}^2}-1 \right]\hat{\mathbf{n}}\cdot \mathbf{v}\nonumber
+\frac{1}{\mathcal{H}}\left(\hat{\mathbf{n}}\cdot\dot{\mathbf{v}}+\frac{\partial\Psi}{\partial r} \right)\nonumber,
\end{align}
where $r$ is the conformal distance of redshift $z$, a dot denotes derivative with respect to conformal time $\eta$ and $\mathcal{H}=\dot a/a$ is the conformal Hubble parameter. $\Delta_\Omega$ is the transverse Laplacian and $L_*$ the luminosity corresponding to $m_*$. Equation~\eqref{eq:numcount} is written in terms of gauge-invariant quantities: $D$ denotes the gauge-invariant density in the synchronous gauge, $\mathbf{v}$ is the gauge-invariant velocity in the longitudinal gauge and $\Phi$ and $\Psi$ are the Bardeen potentials~\footnote{Our convention for the Bardeen potentials is such that in the longitudinal gauge the metric reads $ds^2=-a^2\big[1+2\Psi \big]d\eta^2+a^2\big[1-2\Phi \big]\delta_{ij}dx^i dx^j$.}. All quantities in Equation~\eqref{eq:numcount} are evaluated at the spacetime coordinates corresponding to the observed redshift and angles in the background FRW spacetime, and we have dropped local monopoles and dipoles. In deriving the above expression we have also assumed the background proper source density is time-independent, which should be a reasonable approximation on the scales of interest. We have assumed that tracers have a scale-independent bias $b(z)$. Finally, $s(m_*,z)$ is the slope of the luminosity function expressed in terms of apparent magnitude counts at the flux limit.

We will sometimes refer to the first two terms in Equation~\eqref{eq:numcount} as the `standard' terms. These represent real-space density fluctuations and redshift-space distortions respectively. The third term in Equation~\eqref{eq:numcount} is the so-called magnification bias.
The second line contains the Doppler contributions. We see that besides the term proportional to $-2/(r\mathcal{H})$, which is present in the original derivation of~\citep{1987MNRAS.227....1K,2008MNRAS.389..292P}, we have various other Doppler contributions generated by the fact that we observe on our past-light cone, as discussed in the introduction. In addition to the Doppler contributions, we have in the second line of~\eqref{eq:numcount} a contribution from 
gravitational redshift, proportional to the gradient of $\Psi$. Assuming that galaxies obey the pressure-free Euler equation, this gravitational redshift contribution can be written in terms of the velocity and the second line of Equation~\eqref{eq:numcount} becomes
\be
\label{eq:numcount_euler}
-\left[5s(m_*,z)+\frac{2-5s(m_*,z)}{r\mathcal{H}}+\frac{\dot{\mathcal{H}}}{\mathcal{H}^2}\right]\hat{\mathbf{n}}\cdot \mathbf{v}.
\ee
These contributions are suppressed relative to standard terms by one factor of $\mathcal{H}/k$, where $k$ is the Fourier wavenumber. Since we restrict ourselves to sub-horizon scales these terms are subdominant, but their dipolar nature means they enter into antisymmetric correlation functions at leading order, as we shall soon see. The contributions neglected in Equation~\eqref{eq:numcount} are further suppressed by two factors of $\mathcal{H}/k$ compared to standard terms and they do not contribute to the antisymmetric correlation function. We neglect them therefore in the following. The magnification bias in the first line of Equation~\eqref{eq:numcount} does in principle contribute to both the symmetric and the antisymmetric correlation function, but as shown in \citep{2014PhRvD..89h3535B} the antisymmetric contribution is subdominant at the redshifts considered in this work and we can neglect it.

In addition to the observed galaxy number counts, we also consider the observed fluctuation in the 21\,cm brightness temperature. An expression for this was derived in~\citep{2013PhRvD..87f4026H}. Dropping local monopoles and dipoles, the dominant terms read
\begin{align}
\label{eq:21count}
\Delta_{21}(\hat{\mathbf{n}},z) &\approx b_{21}(z) D - \frac{1}{\mathcal{H}}\hat{\mathbf{n}} \cdot \frac{ \partial \mathbf{v}}{\partial r}
-\left(1+\frac{\dot{\mathcal{H}}}{\mathcal{H}^2} \right)\hat{\mathbf{n}}\cdot \mathbf{v}
+\frac{1}{\mathcal{H}}\left(\hat{\mathbf{n}}\cdot\dot{\mathbf{v}}+\frac{\partial\Psi}{\partial r} \right)
\end{align}
where again all first-order quantities are evaluated on the unperturbed lightcone, and we have neglected terms of order $(\mathcal{H}/k)^2$. Assuming that galaxies obey the pressure-free Euler equation, the Doppler and gravitational redshift contributions in Equation~\eqref{eq:21count} reduce to
\be
\label{eq:21count_euler}
-\left(2+\frac{\dot{\mathcal{H}}}{\mathcal{H}^2}\right)\hat{\mathbf{n}}\cdot \mathbf{v}\, .
\ee
It may be noticed that the fluctuations in the brightness temperature could have been derived from Equation~\eqref{eq:numcount} by setting $s(m_*,z) = 2/5$. The reason for this is twofold. Firstly, the linear fluctuation in the number counts receives contributions from sources being magnified over the flux limit of the survey. This effect is not present in the 21\,cm intensity map, as effectively all the photons in the frequency channel contribute to the observable, and individual objects are not resolved. Secondly, notwithstanding the magnification effect, linear fluctuations in brightness temperature are related to those in number counts by the perturbation to the luminosity distance~\citep{2013PhRvD..87f4026H,Bonvin:2005ps,Hui:2005nm}. These cancel terms in the number count perturbation that arise from the transformation of the source volume element to the observed volume element in redshift/solid angle coordinates, in particular the lensing convergence. Thus, 21\,cm intensity maps do not experience the boosting of observed fluctuations from magnification, nor the suppression of fluctuations from the expansion of the source volume element. By choosing $s(m_*,z) = 2/5$ at every redshift, both effects may be removed via exact cancellation.

We will define the redshift-space cross-correlation function by $\langle \Delta_A(\mathbf{r};z)\Delta_B(\mathbf{r'};z')\rangle \equiv \xi^{AB}(\mathbf{d};z,z')$ with $\mathbf{d}=\mathbf{r'}-\mathbf{r}$, which assumes statistical homogeneity. The population indices $(A,B)$ could refer either to 21\,cm intensity maps, or individual populations of galaxies within a number counts survey. For example, it may be possible to split the sample into bright and faint sources, or split by galaxy type. We may further define the correlation function multipoles by expanding in Legendre polynomials $\mathcal{L}_\ell$ as
\begin{equation}
\label{eq:mult}
\xi^{AB}(\mathbf{d};z,z') = \sum_{\ell=0}^4 \xi^{AB}_{\ell}(d;z,z') \mathcal{L}_\ell(\hat{\mathbf{d}} \cdot \hat{\mathbf{n}}),
\end{equation}
where $\hat{\mathbf{n}}$ denotes the direction of the pair~\footnote{Note that at large separation, wide-angle corrections generate multipoles higher than $\ell=4$.}. As shown in~\cite{2015arXiv151203918G}, one can choose different definitions for $\hat{\mathbf{n}}$: it can either be the direction of the midpoint of the pair-separation vector $\mathbf{d}$, or it can be the direction of either one of the member of the pair. In the distant-observer limit, all definitions are equivalent. However, as we shall see in section~\ref{sec:wideangle}, the Doppler signal is contaminated by wide-angle effects. These wide-angle effects depend closely on the definition of $\hat{\mathbf{n}}$ and, as shown in~\cite{2015arXiv151203918G}, they are minimal when $\hat{\mathbf{n}}$ is the direction of the midpoint of the pair-separation vector $\mathbf{d}$. We will therefore adopt this definition in the rest of this work.
 
To make progress, we take the product of Equations~\eqref{eq:numcount} and \eqref{eq:21count}, and expand first-order terms in Fourier modes, with the convention
\begin{equation}
\Delta(\mathbf{r}) = \int \frac{\mathrm{d}^3\mathbf{k}}{(2\pi)^3}\, \Delta(\mathbf{k}) \mathrm{e}^{i \mathbf{k} \cdot \mathbf{r}}.
\end{equation}

\subsection{Distant-observer limit}

In the distant observer limit where the line-of-sight vector $\hat{\mathbf{n}}$ is assumed fixed, the correlation multipoles may be related to those of the power spectrum. With our Fourier transform convention the relationship is
\begin{equation}
\xi^{AB}_{\ell}(d;z,z') = (-i)^{\ell} \int \frac{k^2 \mathrm{d}k}{2 \pi^2} \, P^{AB}_{\ell}(k;z,z') j_\ell(kd),
\end{equation}
where the power spectrum is defined as $\langle \Delta_A(\mathbf{k};z) \Delta_B(\mathbf{k}';z') \rangle = (2\pi)^3 P^{AB}(\mathbf{k};z,z') \delta^D(\mathbf{k} + \mathbf{k}')$, and the power spectrum multipoles are defined in analogy to the correlation function multipoles. The odd power spectrum multipoles are derived in~\citep{2009JCAP...11..026M, 2014PhRvD..89h3535B,Raccanelli:2013dza}, whilst the even multipoles follow from the standard Kaiser redshift-space distortion formula~\citep{1987MNRAS.227....1K}. The full set of multipoles at leading order in $\mathcal{H}/k$ are given by
\begin{align}
P^{AB}_0(k) &= \left[ b_Ab_B + \frac{f}{3}(b_A + b_B) + \frac{f^2}{5}\right]P_m(k) \nonumber \\
P^{AB}_1(k) &= (-i)\left\{\left(\frac{2}{r\mathcal{H}} + \frac{\dot{\mathcal{H}}}{\mathcal{H}^2}\right)(b_B - b_A) + \left(1 - \frac{1}{r\mathcal{H}}\right)\left[3f(s_A - s_B) + 5(b_Bs_A - b_As_B)\right]\right\}  f \frac{\mathcal{H}}{k} P_m(k) \nonumber \\
P^{AB}_2(k) &=\left[\frac{2f}{3}(b_A + b_B) + \frac{4f^2}{7}\right]P_m(k) \nonumber \\
P^{AB}_3(k) &= 2i\left(1 - \frac{1}{r\mathcal{H}}\right)(s_B - s_A)f^2P_m(k)\frac{\mathcal{H}}{k}\nonumber \\
P^{AB}_4(k) &= \frac{8f^2}{35}P_m(k),
\label{eq:pmults}
\end{align}
where $P_m(k)$ is the matter power spectrum, and the time-dependence has been left implicit. In deriving these expressions, we have used the Euler equation and we have further assumed that velocities are related to the density field by a scale-independent growth factor $f(z)$. Thus, it is clear that odd multipoles directly probe large-scale bulk flows, both through their auto-correlation and cross-correlation with the matter density. Specifically, the dipole receives contributions from cross-correlations of Doppler terms in Equations~\eqref{eq:numcount_euler} and~\eqref{eq:21count_euler} with density terms, and from cross-correlations of Doppler terms with redshift-space distortion terms. The octupole has contributions only from cross-correlations of Doppler and redshift-space distortions. Thus, the odd moments provide complementary information on bulk-flows to that provided by the even moments. Although both even and odd moments are sensitive to the same velocity field, the odd moments probe a physically distinct effect: as we can see from Equation~\eqref{eq:numcount} the odd moments follow from the impact of the peculiar velocity on the observed position of the pixel, whereas the even moments are generated by a change in the observed size of the pixel. Furthermore, the differing scale-dependence of the Doppler and redshift-space distortion terms means the odd and even multipoles are not completely correlated, and hence the odd multipoles contain independent information on large-scale flows. As well as improving error bars on cosmological parameters affecting structure growth, any disparity in the amplitude of $f\sigma_8$ measured from odd moments compared to that measured from even moments could be indicative of a breakdown of the equivalence principle, which has been assumed in deriving Equations~\eqref{eq:numcount_euler} and~\eqref{eq:21count_euler}, or the presence of scale-dependent growth.

\begin{figure}
\centering
\includegraphics[width=0.6\textwidth]{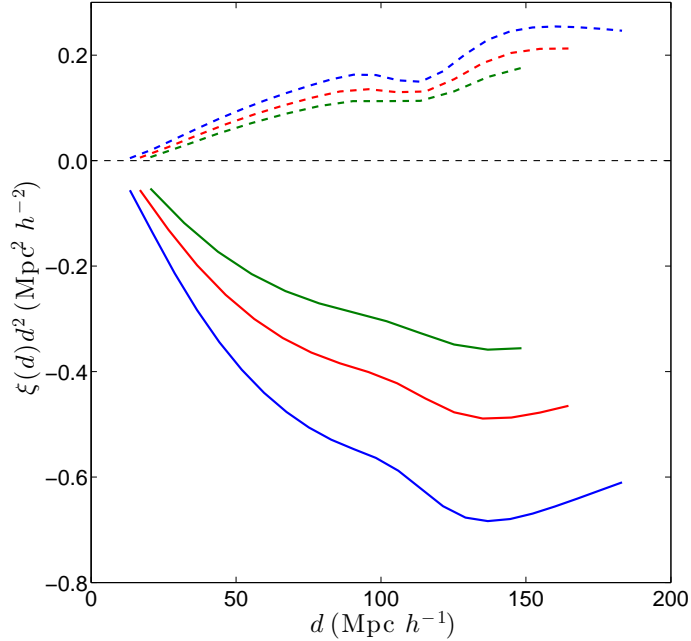}
\caption{(Colour Online). Cross-correlation dipole between CHIME 21\,cm intensity emission and Euclid ELGs as a function of the pair separation $d$. The signal is plotted at (from bottom to top) $z=1.0$ (blue solid), $z=1.2$ (red solid), and $z=1.4$ (green solid). We also plot the wide-angle terms at (from top to bottom) $z=1.0$ (blue dashed), $z=1.2$ (red dashed), and $z=1.4$ (green dashed). The black dashed line indicates zero cross-correlation. The minimum and maximum values of $d$ shown for these curves are the those used in the $S/N$ calculations at each redshift.}
\label{fig:CHIME_EuclidELG_SIG}
\end{figure}

In Figure~\ref{fig:CHIME_EuclidELG_SIG} we plot the redshift-space cross-correlation dipole for a typical 21\,cm and galaxy survey that will be used in this work. The sign of the dipole is determined by the relative biases and magnification slopes, and its redshift dependence depends additionally on the growth of structure through the matter power spectrum. Also apparent is the peak due to baryon acoustic oscillations, visible in the dipole as it is in the more conventional even multipole moments.

In this work we will focus primarily on the dipole of the correlation function. The octupole also contains information on the large-scale growth of structure, but is only non-zero if the magnification slopes of the two surveys differ. Since the forecast values of these slopes carry much more uncertainty than the biases, we will focus only on the dipole, noting that the signal-to-noise could be boosted further with inclusion of information from the octupole.

\subsection{Wide-angle corrections}
\label{sec:wideangle}

The dipole of the cross-correlation function is proportional to $\mathcal{H}/k$, and so it may be worried that corrections to the standard even moments due to departures from the distant-observer limit may be of comparable magnitude. The leading order corrections are proportional to $d/r$ and their specific form depends on the definition of the angle $\hat{\mathbf{d}} \cdot \hat{\mathbf{n}}$ used to measure the multipoles (see~Equation~\eqref{eq:mult})~\citep{2015arXiv151203918G,Reimberg:2015jma}. Choosing $\hat{\mathbf{n}}$ as the direction of the midpoint of the pair-separation vector $\mathbf{d}$ minimises the wide-angle (WA) corrections, which become~\cite{2014PhRvD..89h3535B,2015arXiv151203918G}
\begin{equation}
\xi^{AB,{\rm WA}}_1(d) = \frac{2f}{5}(b_B-b_A) \frac{d}{r} \int \frac{k^2\mathrm{d}k}{2\pi^2}\, P_m(k)j_2(kd).
\label{eq:wasig}
\end{equation}
This expression neglects contributions from evolution in the biases and growth factor, and is thus only valid on scales smaller than the characteristic time-scales for this evolution. For the bias models under consideration in this work, it was shown in~\cite{2014PhRvD..89h3535B} that Equation~\eqref{eq:wasig} should be the leading order contribution to wide-angle effects.

In Figure~\ref{fig:CHIME_EuclidELG_SIG} we plot the wide-angle term alongside the dipole signal. The two are clearly of comparable magnitude, so for most of the models and surveys we consider in this work it will be necessary to correct the measured dipole for this contaminating term. Alternatively, the wide-angle term may be considered an extra source of signal - being proportional to $f$ it also probes large-scale flows. However, this dependence arises from cross-correlations of density and redshift-space distortions, and so does not provide new information to that contained within the even moments. Since we are concerned only with signal arising from the Doppler terms in Equations~\eqref{eq:numcount} and~\eqref{eq:21count}, we consider the wide-angle term a contaminant to be removed.

As shown in~\citep{2014PhRvD..89h3535B}, removal of the wide-angle term may be achieved by modifying the dipole estimator as
\begin{equation}
\hat{\xi}^{AB}_1 \rightarrow \hat{\xi}^{AB}_1 - \frac{3}{10}\left(\hat{\xi}^{BB}_2 - \hat{\xi}^{AA}_2\right)\frac{d}{r}.
\label{eq:waest}
\end{equation}
This modified estimator is now unbiased by wide-angle effects, although some extra variance is incurred which we calculate in the next section\footnote{In this work we focus on frequentist point-estimators for the dipole correlation function. Alternatively we could package all the observables into a data vector, model the wide-angle bias, and construct a likelihood function for the true correlation function multipoles. Although this Bayesian approach is optimal, the functional form of the likelihood is unknown in the non-linear regime, and an incorrect specification could bias parameter inferences. We avoid this problem in our approach - even though we have assumed Gaussian statistics for our estimators this could easily be relaxed and the variances estimated from simulations. In contrast, estimating the full multivariate likelihood function from simulations is extremely challenging.}. Note that throughout this work we will only consider galaxy surveys with spectroscopic redshifts available, and thus $r$ and $d$ are both assumed to be known perfectly.

\section{Estimators and covariance matrix}
\label{sec:var}

In this section we write down our estimator for the cross-correlation dipole and compute its covariance matrix. The approach here is similar to that of~\citep{2016JCAP...08..021B}, except we provide a much more compact formula for computing the covariance matrix, valid in the distant-observer limit and assuming Gaussianity. This latter assumption is valid on large-scales, but breaks down on small scales due to non-linear clustering and non-linear redshift-space distortions, a point we return to in Section~\ref{subsec:systematics}.

\subsection{Distant-observer limit}

Neglecting wide-angle effects and taking the continuum-limit, our estimator for the orientation-dependent redshift-space multi-tracer correlation function is
\begin{equation}
\hat{\xi}^{AB}(\mathbf{d}) = \int \frac{\mathrm{d}^3\mathbf{r}}{V}\, \hat{\Delta}_A(\mathbf{r} - \mathbf{d}/2) \hat{\Delta}_B(\mathbf{r} + \mathbf{d}/2),
\label{eq:est}
\end{equation}
where $V$ is the volume of the survey, and $\hat{\Delta}$ is the measured fractional overdensity in either the number counts or 21\,cm brightness temperature in a pixel. We will assume that the background 21\,cm monopole can either be measured and removed using the auto-spectrum, or alternatively can be marginalised over in a likelihood for the cross-correlation function conditioned on the auto-spectrum. Our estimator is biased by Poisson noise only when $A=B$ and $\mathbf{d}=0$. Note that this estimator ensures that swapping $A\leftrightarrow B$ is equivalent to flipping the sign of $\mathbf{d}$. Note also that we have neglected effects such as the survey window function and inhomogeneous noise. A more optimal estimator could account for these with a weight function $w(\mathbf{r})$ in Equation~\eqref{eq:est}.

Taking the two-point function of Equation~\eqref{eq:est} and using that~\citep{2009MNRAS.400..851S}
\begin{equation}
\langle \hat{\Delta}_A(\mathbf{r}) \hat{\Delta}_B(\mathbf{r}') \rangle = \xi^{AB}(\mathbf{r}' - \mathbf{r}) + \frac{\delta^K_{AB}}{\bar{n}_A}\delta^D(\mathbf{r}'-\mathbf{r}),
\end{equation}
where $\bar n_A$ is the number density of population $A$, we obtain the covariance matrix
\begin{align}
C^{AB}_{CD}(\mathbf{d}, \mathbf{d}') &= \int \frac{\mathrm{d}^3\mathbf{r}}{V} \int \frac{\mathrm{d}^3\mathbf{r}'}{V} \left[\vphantom{\frac12} \xi^{AC}(\mathbf{r}' - \mathbf{d}'/2 - \mathbf{r} + \mathbf{d}/2) \xi^{BD}(\mathbf{r}' + \mathbf{d}'/2 - \mathbf{r} - \mathbf{d}/2) \nonumber \right.\\
&\quad \left. {} + \xi^{AD}(\mathbf{r}' + \mathbf{d}'/2 - \mathbf{r} + \mathbf{d}/2)\xi^{BC}(\mathbf{r}' - \mathbf{d}'/2 - \mathbf{r} - \mathbf{d}/2)  \vphantom{\frac12}\right] \nonumber \\
 & +  \xi^{BD}(\mathbf{d}' - \mathbf{d})\frac{\delta^K_{AC}}{\bar{n}_AV} +  \xi^{AC}(\mathbf{d} - \mathbf{d}')\frac{\delta^K_{BD}}{\bar{n}_BV} +  \xi^{AD}(\mathbf{d}' + \mathbf{d})\frac{\delta^K_{BC}}{\bar{n}_BV} +  \xi^{BC}(-\mathbf{d}' - \mathbf{d})\frac{\delta^K_{AD}}{\bar{n}_AV}  \nonumber \\
& + \delta^D(\mathbf{d}' - \mathbf{d}) \frac{\delta^K_{AC}\delta^K_{BD}}{\bar{n}_A \bar{n}_B V} +  \delta^D(\mathbf{d}' + \mathbf{d}) \frac{\delta^K_{BC}\delta^K_{AD}}{\bar{n}_A \bar{n}_B V},
\end{align}
which assumes Gaussianity. We shall refer to the first two lines of this expression as the Cosmic Variance $\times$ Cosmic Variance term, the third line as the Cosmic Variance $\times$ Poisson term, and the final line as the Poisson $\times$ Poisson term. Note that in the case where one of the tracers is the 21\,cm brightness temperature there will also be a contribution from 21\,cm interferometer noise in the correlation function. In that case we will also have Cosmic Variance $\times$ Interferometer Noise, Poisson $\times$ Interferometer Noise, and potentially Interferometer Noise $\times$ Interferometer Noise. Later on we will discuss the form of these contributions.

One of the integrals in the Cosmic Variance $\times$ Cosmic Variance term may be done trivially by changing variables to $\mathbf{r}'' = \mathbf{r}' - \mathbf{r}$. Following~\citep{2001ApJ...546....2E} we expand the correlation function in Fourier modes, which allows the $\mathbf{r}$-integral to be done exactly, yielding a Dirac delta-function which leaves the result as
\begin{equation}
\frac{1}{V}\int \frac{\mathrm{d}^3\mathbf{k}}{(2\pi)^3} \, \left[P_{AC}(k,\mu)P_{BD}(k,-\mu)\mathrm{e}^{-i\mathbf{k}\cdot (\mathbf{d}-\mathbf{d}')} + P_{AD}(k,\mu)P_{BC}(k,-\mu) \mathrm{e}^{-i \mathbf{k} \cdot (\mathbf{d} + \mathbf{d}')} \right],
\label{eq:cvcv}
\end{equation}
where recall that $\mu$ is the cosine of the angle between the wavevector and the line-of-sight, and by definition $P_{AB}(k,\mu) = P_{BA}(k,-\mu)$.

We now integrate the estimator against an orientation-dependent weight function $w(\hat{\mathbf{d}})$, which for the dipole would just be the $\ell=1$ Legendre polynomial. Using the result
\begin{equation}
\int \frac{\mathrm{d}^2\Omega_{\hat{\mathbf{d}}}}{4\pi} w(\hat{\mathbf{d}}) \mathrm{e}^{-i \mathbf{k} \cdot \mathbf{d}} = \sum_{\ell=0}^{\infty} i^{-\ell} \mathcal{L}_\ell(\mu) w_\ell j_\ell(kd),
\end{equation}
where $w_\ell$ are the Legendre moments of the weight function, we have for the Cosmic Variance $\times$ Cosmic Variance term
\begin{align}
&\frac{1}{V}\int \frac{\mathrm{d}^3\mathbf{k}}{(2\pi)^3} \, \left[P_{AC}(k,\mu)P_{BD}(k,-\mu) \sum_{\ell,\ell'} i^{\ell'-\ell}\mathcal{L}_\ell(\mu)\mathcal{L}_{\ell'}(\mu)w_\ell w_{\ell'} j_\ell(kd) j_{\ell'}(kd') \nonumber \right.\\
& \quad \left. {} + P_{AD}(k,\mu)P_{BC}(k,-\mu) \sum_{\ell,\ell'} i^{\ell'+\ell}\mathcal{L}_\ell(\mu)\mathcal{L}_{\ell'}(\mu)w_\ell w_{\ell'} j_\ell(kd) j_{\ell'}(kd') \vphantom{\frac12}\right].
\end{align}
We now expand the power spectra in Legendre polynomials, and perform the angular integral. Following similar steps for the Poisson terms and identifying $\delta^D(d-d') = \delta^K_{d,d'}/L_p$ in the discrete limit with square pixels of side-length $L_p$, we have our final result for the covariance matrix
\begin{align}
&C_{AB}^{CD}(d,d') = \frac{1}{V} \int \frac{k^2 \mathrm{d}k}{2\pi^2}\, \sum_{\ell,\ell'} i^{\ell'-\ell} w_\ell w_{\ell'} j_\ell(kd) j_\ell(kd') \sum_{L,L'} G_{\ell'\ell}^{L'L} \left[P^{AC}_L(k)P^{DB}_{L'}(k) + (-1)^{\ell'}P^{AD}_L(k)P^{CB}_{L'}(k)\right] \nonumber \\
& + \int \frac{k^2 \mathrm{d}k}{2\pi^2}\,  \sum_{\ell,\ell'}i^{\ell'-\ell}w_\ell w_{\ell'} j_\ell(kd) j_\ell(kd') \sum_{L} \ThreeJSymbol{L}{0}{\ell}{0}{\ell'}{0}^2\left[\frac{\delta^K_{AC}}{\bar{n}_AV}P^{DB}_{L}(k) + \frac{\delta^K_{BD}}{\bar{n}_BV}P^{CA}_{L}(k) \nonumber \right. \\
& \quad \left. {} + (-1)^{\ell'}\frac{\delta^K_{AD}}{\bar{n}_AV}P^{BC}_{L}(k) + (-1)^{\ell'}\frac{\delta^K_{BC}}{\bar{n}_BV}P^{DA}_{L}(k)  \vphantom{\frac12}\right] \nonumber \\
&+ \frac{\delta^K_{AC}\delta^K_{BD}}{\bar{n}_A\bar{n}_BV} \frac{\delta^{K}_{d,d'}}{4\pi d^2L_p}\sum_\ell\frac{w_\ell^2}{2\ell+1} + \frac{\delta^K_{BC}\delta^K_{AD}}{\bar{n}_A\bar{n}_BV} \frac{\delta^{K}_{d,d'}}{4\pi d^2L_p}\sum_\ell (-1)^\ell\frac{w_\ell^2}{2\ell+1}.
\label{eq:covmat}
\end{align}
Note that we could have packaged up the Poisson terms with the power spectra to leave an expression similar to first line of the above equation, but the form presented here is more useful for computing the individual contributions to the estimator noise. The quantity $G_{\ell'\ell}^{L'L}$ arising from the integral of four Legendre polynomials is expressible in terms of Wigner 3$j$ symbols as
\begin{equation}
G^{L' L}_{\ell' \ell} \equiv \sum_{L''}(2L''+1) \ThreeJSymbol{\ell}{0}{\ell'}{0}{L''}{0}^2 \ThreeJSymbol{L}{0}{L'}{0}{L''}{0}^2,
\end{equation}
which obeys the symmetries $G_{\ell'\ell}^{L'L} = G_{\ell\ell'}^{L'L} = G_{\ell'\ell}^{LL'} = G_{L'L}^{\ell'\ell}$ and is only non-zero when the parity of $\ell+\ell'$ equals that of $L+L'$. Note also that $P^{AB}_\ell(k) = (-1)^\ell P^{BA}_\ell(k)$. An expression similar to Equation~\eqref{eq:covmat} recently appeared in~\citep{2016MNRAS.457.1577G}; our result reduces to theirs in the single-population case.

We have thus succeeded in reducing a nine-dimensional integral to a finite sum of one-dimensional integrals, which can be computed rapidly. The above expression generalises previous formulae for the Gaussian covariance matrix~\citep[e.g.][]{1994ApJ...424..569B, 2006NewA...11..226C, 2009MNRAS.400..851S, 2016MNRAS.457.1577G} to the multi-tracer redshift-space case with arbitrary orientation weights.

As found in~\citep{2016JCAP...08..021B}, the standard density and redshift-space distortions {\it do not contribute} to the Cosmic Variance $\times$ Cosmic Variance term for odd weights. This is in fact immediately obvious from Equation~\eqref{eq:cvcv}. Substituting the Kaiser formula for the power spectra, we see that the two products of power spectra in Equation~\eqref{eq:cvcv} are equal to each other. The resulting term then vanishes when integrated against an odd weight function. This shows that our estimator allows us to get rid of the dominant density and redshift-space distortions not only in the signal but also the cosmic variance. As a consequence, the Cosmic Variance $\times$ Cosmic Variance term is generated only by the subdominant Doppler terms.

Since we are only concerned with the dipole in this work, we will set $w_\ell = 3\delta^K_{\ell,1}$, where the normalisation ensures an unbiased result. This choice, combined with the various symmetries in the Wigner 3$j$ symbols, ensures $C_{AB}^{CD}(d,d') = - C_{BA}^{CD}(d,d') = -C_{AB}^{DC}(d,d')$. We also have $C_{AB}^{CD}(d,d') = C_{CD}^{AB}(d',d)$, as required by symmetry.

\begin{figure}
\centering
\includegraphics[width=0.45\textwidth]{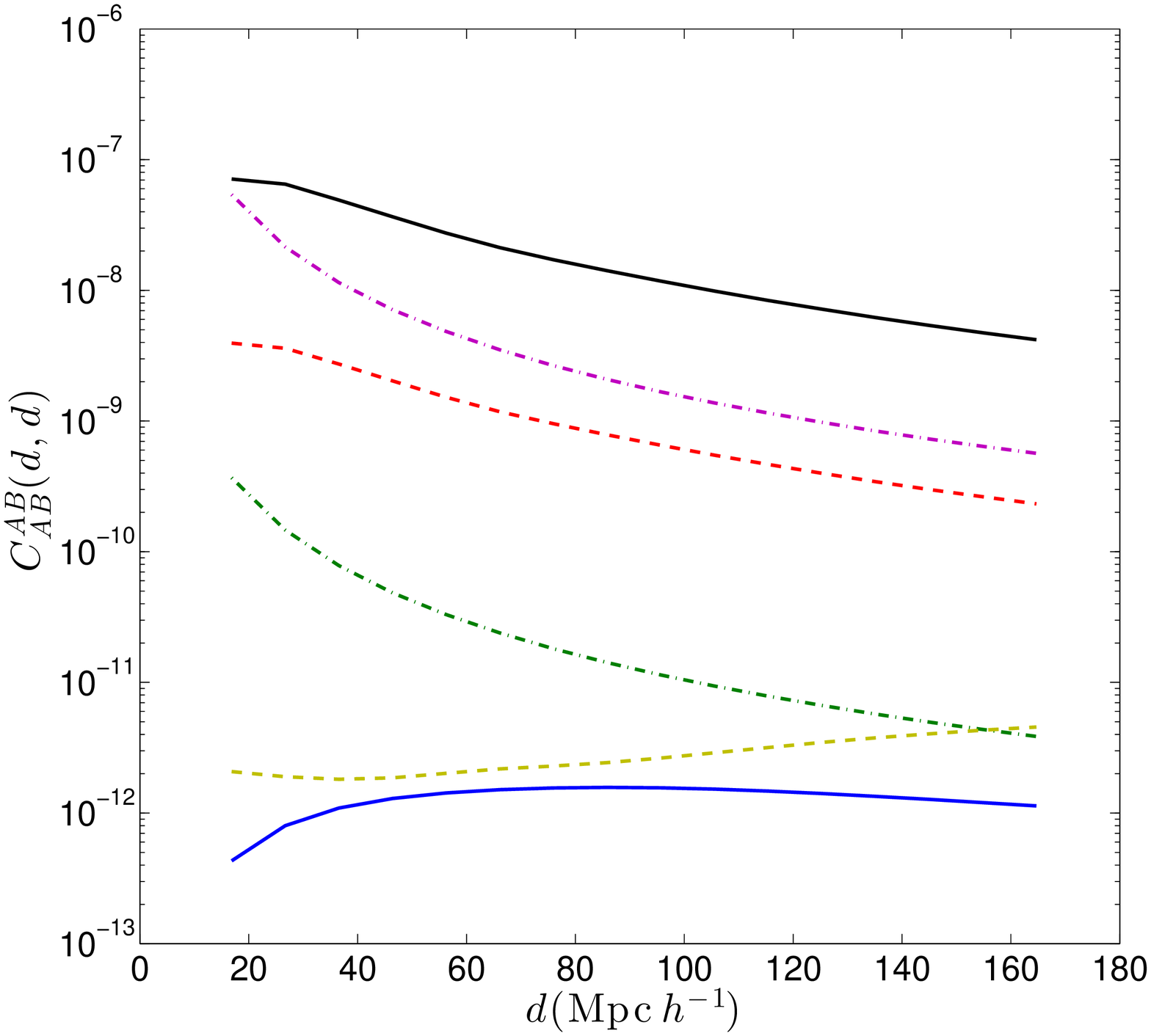}
\includegraphics[width=0.45\textwidth]{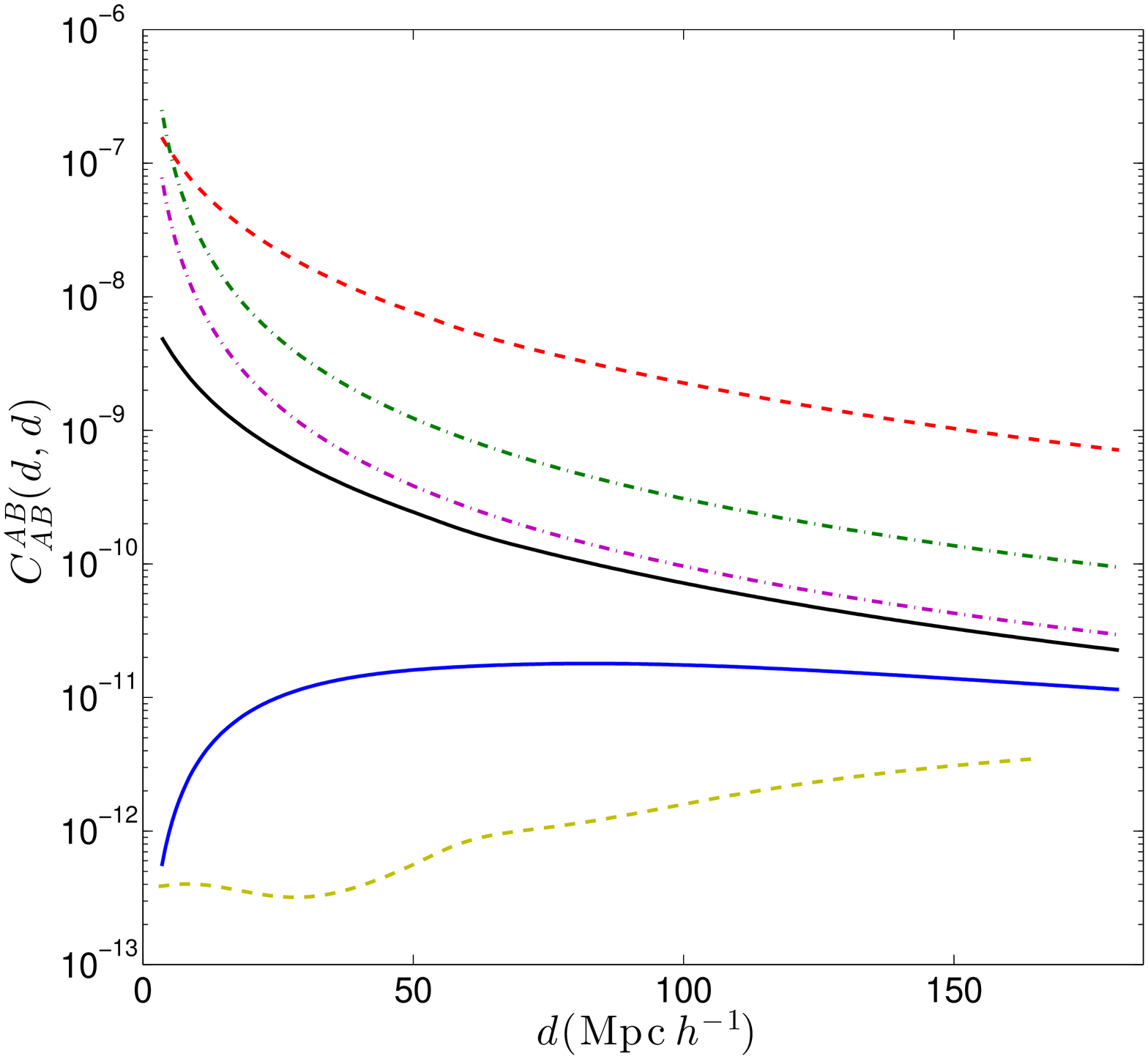}
\caption{(Colour Online). Contributions to the variance of the cross-correlation dipole between CHIME (\emph{left panel}) and SKA (\emph{right panel}) with Euclid ELGs at $z=1.2$. Plotted are the contributions from Cosmic Variance $\times$ Cosmic Variance (blue lower solid), Cosmic Variance $\times$ Poisson (red upper dashed), Cosmic Variance $\times$ Interferometer Noise (black upper solid), Poisson $\times$ Interferometer Noise (magenta upper dot-dashed in left panel, lower dot-dashed in right panel), Poisson $\times$ Poisson (green lower dot-dashed in left panel, upper dot-dashed in right panel), and the extra variance incurred by removing the wide-angle term (yellow lower dashed). The pixel size is $L_p = 9.9 \, \mathrm{Mpc}/h$ for CHIME and $L_p = 1.3 \, \mathrm{Mpc}/h$ for SKA.}
\label{fig:CHIME_SKA_EuclidELG_cov}
\end{figure}

In Figure~\ref{fig:CHIME_SKA_EuclidELG_cov} we plot the contributions of the Cosmic Variance $\times$ Cosmic Variance, the Cosmic Variance $\times$ Poisson and the Poisson $\times$ Poisson terms to the diagonal elements of the covariance matrix, for correlations between a number counts survey (Euclid emission line galaxies) and two different intensity mapping surveys having high angular resolution (SKA) and low angular resolution (CHIME) at $z=1.2$. In both cases we see that the Cosmic Variance $\times$ Cosmic Variance contribution to the variance is negligible. Its amplitude is higher for SKA due to the higher amplitude of the beam-smoothed dipole power spectrum due to the smaller smoothing scale used as a consequence of SKA's higher resolution (see Section~\ref{sec:21noise}). The Cosmic Variance $\times$ Poisson term is the dominant term for SKA but subdominant for CHIME, again due to the higher amplitude of the beam-smoothed power spectrum for SKA.  Finally, the Poisson $\times$ Poisson contribution is negligible for CHIME and subdominant for most separations for SKA. The SKA has a higher resolution and hence a smaller $L_p$, and thus higher Poisson noise to the presence of this factor in the denominators of the last line of Equation~\eqref{eq:covmat}.

\begin{figure}
\centering
\includegraphics[width=0.45\textwidth]{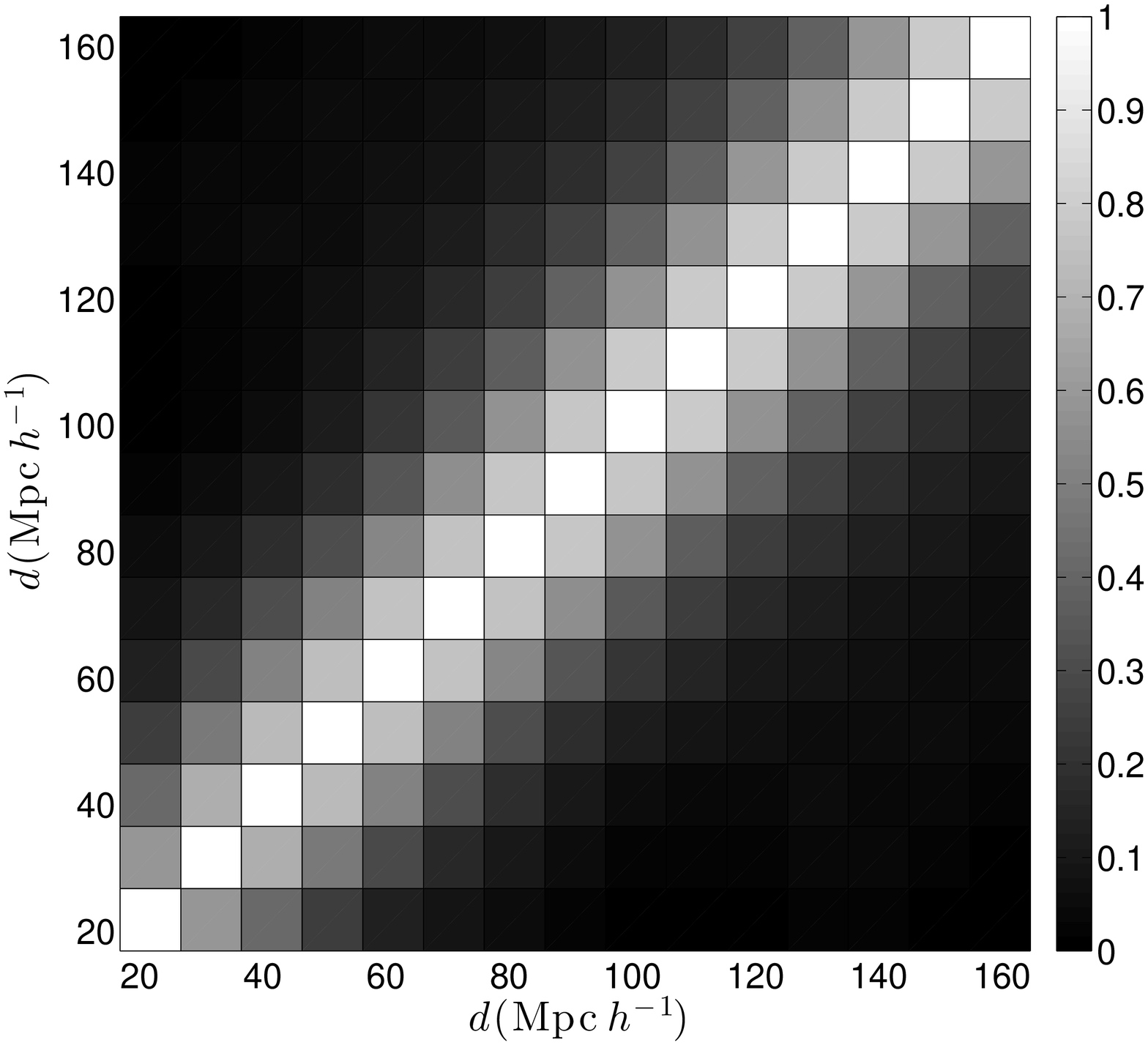}
\includegraphics[width=0.45\textwidth]{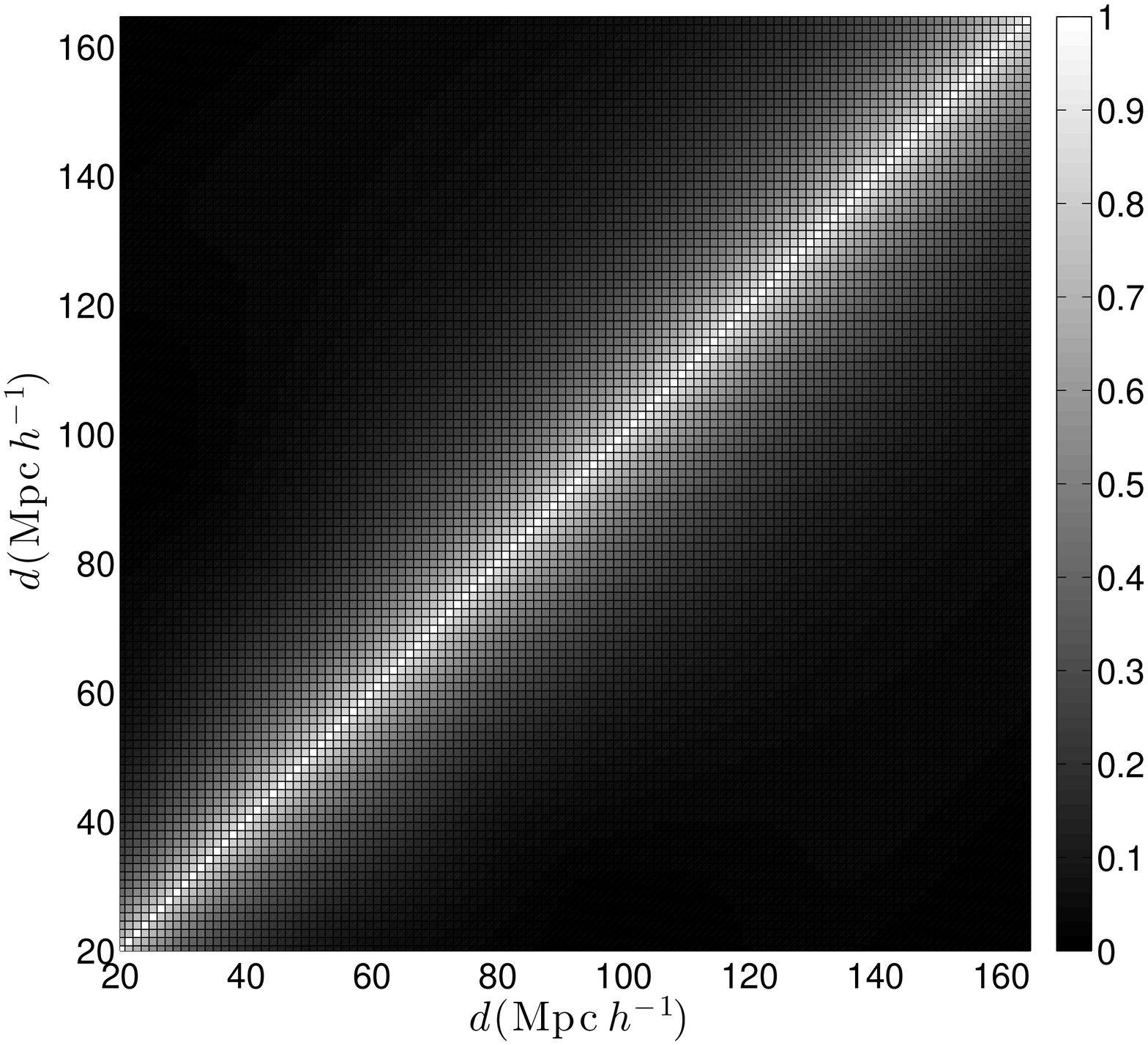}
\caption{(Colour Online). Absolute value of the correlation matrix for the cross-correlation dipole between CHIME (\emph{left panel}) and SKA (\emph{right panel}) with Euclid ELGs at $z=1.2$.}
\label{fig:CHIME_SKA_EuclidELG_corr}
\end{figure}

In Figure~\ref{fig:CHIME_SKA_EuclidELG_corr} we plot the dimensionless correlation matrix of the dipole-dipole two-population estimator. Firstly we see that the SKA has much higher resolution than CHIME, due to the larger area of its array. We also see that there are off-diagonal correlations induced due to cosmic variance, with a typical correlation length being roughly $40 \, \mathrm{Mpc}\,h^{-1}$ for CHIME and $20 \, \mathrm{Mpc}\,h^{-1}$ for SKA. The smaller correlation length for the SKA is likely due to the relatively greater importance of Poisson noise to the dipole variance (see Figure~\ref{fig:CHIME_SKA_EuclidELG_cov}).

\subsection{Wide-angle corrections}

We have seen in Section~\ref{sec:sig} that the contribution to the signal from wide-angle terms is significant, which motivates the estimator proposed in Equation~\eqref{eq:waest}. Extra variance is incurred as a cost, both from correlations between the dipole estimator and the quadrupole correction at order $(d/r)\times(\mathcal{H}/k)$ and the auto-correlation of the quadrupole correction at order $(d/r)^2$. These are of similar order, and turn out to partially cancel each other for most of the bias models and surveys we consider, which makes the wide-angle variance effectively negligible. Nevertheless we include both contributions for completeness.

The wide-angle contributions may be straightforwardly computed from Equation~\eqref{eq:covmat}, with the quadrupole weights given by $w_\ell = 5\delta^K_{\ell,2}$. Note that Poisson $\times$ Poisson, Poisson $\times$ Interferometer Noise, and Interferometer Noise $\times$ Interferometer Noise do not contribute to the $(d/r)\times(\mathcal{H}/k)$ cross-term due to symmetry.

In Figure~\ref{fig:CHIME_SKA_EuclidELG_cov} we plot the wide-angle contribution to the dipole variance. It is clear from this plot that the contribution is negligible even for the largest separations considered for these surveys, due to a combination of $(\mathcal{H}/k)$ suppression, $(d/r)$ suppression, and the fortuitous cancellation described above. This shows that the estimator~\eqref{eq:waest} provides a robust way of isolating the Doppler contributions.

\subsection{Interferometer noise and beams}
\label{sec:21noise}

In addition to Poisson noise and cosmic variance, 21\,cm intensity maps contain contributions from both foregrounds and noise from the interferometer\footnote{We do not consider single-dish experiments in this work - doing so would require only a straightforward modification to the noise power spectrum~\citep{2015ApJ...803...21B}}. We will assume that the 21\,cm intensity maps have been cleaned for foregrounds, and we will neglect their contribution to the variance. We note however that one of the advantages of using cross-correlations is that foregrounds will not bias the signal as they have a non-cosmological origin. A more detailed assessment of the impact of residual foregrounds on the dipole variance would require modelling their power spectra and spectral properties, as well as the scanning strategy and polarization leakage of the interferometer. We defer investigation of these effects to a future work.

Interferometer noise enters only into the covariance matrix, through its contribution to the power spectrum. This receives an additional contribution $P_N$, which we will be assumed scale-independent. We will also assume that the noise is isotropic, and hence only contributes to the monopole of the power spectrum. This is a common assumption in the literature~\citep[e.g.][]{2010ApJ...721..164S}, but is likely incorrect due to the anisotropic baseline coverage of the interferometer~\citep[e.g.][]{2014ApJ...788..106P}. We will hence assume every baseline is observed for the same amount of time, noting that corrections to this will modify our noise expressions, but will not bias our estimator. 

The 21\,cm brightness temperature itself is modulated by the beam in the angular direction, and by the frequency response in the frequency direction. We thus smooth the map with an isotropic square top-hat function having the same side-length as the angular resolution. Note that we beam-smooth the signal rather than the noise - the noise is assumed to be added after the effects of the beam have been included\footnote{This is not quite correct for an interferometer, since the beam enters the visibility through a Fourier-space convolution rather than a real-space convolution. Our calculation would be appropriate for a single-beam experiment~\citep{1996MNRAS.281.1297T}, but is only an approximate treatment of beams in the case of an interferometer. In any case, some smoothing at map-level will be required to mitigate the influence of non-linear power in the correlation function.}. 

Following~\citep{2015ApJ...803...21B}, we write the three-dimensional noise power spectrum of the fractional overdensity of the 21\,cm brightness temperature as
\begin{equation}
P_N(z) = \frac{T_{\mathrm{sys}}(z)^2}{T_0(z)^2} X(z)^2Y(z) \frac{\lambda(z)^4}{A_{\mathrm{eff}}^2}\frac{S_{21}}{\mathrm{FOV}(z)}\frac{1}{n(\mathbf{u};z)n_pt_0},
\end{equation}
where the system temperature $T_{\mathrm{sys}}$ is related to the antenna temperature $T_{\mathrm{ant}}$ by $T_{\mathrm{sys}}= T_{\mathrm{ant}} + T_{\mathrm{sky}}$, with the sky temperature fixed to $T_{\mathrm{sky}} = 60(300 \mathrm{MHz}/\nu)^{2.55}\mathrm{K}$. $A_{\mathrm{eff}}$ is the effective beam area, whilst $\lambda$ and $\nu$ are the observed wavelength and frequency corresponding to the observed redshift. $S_{21}$ is the survey area in steradians, with $\mathrm{FOV}$ the field-of-view available to each element, assumed to be isotropic. $n_p$ is the number of polarization states, $t_0$ is the total observing time, and $n(\mathbf{u})$ is the number density of baselines in visibility space, which we assume to be uniform as described above, with $n(\mathbf{u}) = N_f^2/2\pi u_{\mathrm{max}}^2$, where $N_f$ is the total number of interferometer elements, and the maximum baseline is given roughly by $D_{\mathrm{max}}/\lambda$ where $D_{\mathrm{max}}$ is the maximum separation of array elements. We have assumed that $N_f \gg 1$, such that $N_f^2/2$ is roughly the number of antenna pairs. 

The quantities $X$ and $Y$ are factors serving to convert between angular and frequency space to physical three-dimensional space. $X$ is the comoving angular diameter distance and $Y = (1+z)^2/[\nu_{21}H(z)]$, where $H(z)$ is the Hubble parameter and $\nu_{21}$ is the frequency of the 21\,cm transition. Finally, the 21\,cm monopole is given by~\citep[e.g.][]{2006PhR...433..181F}
\begin{equation}
T_0(z) = \frac{3}{32\pi}\frac{ h_p A_{10}}{k_B m_p} \frac{\lambda_{21}^2(1+z)^2}{H(z)}\Omega_{\mathrm{HI}}\rho_c,
\end{equation}
where $h_p$ is Planck's constant, $k_B$ is Boltzmann's constant, $m_p$ the mass of the proton, $A_{10} \approx 2.869\times 10^{-15} \, \mathrm{s}^{-1}$ is the 21\,cm spontaneous emission coefficient, $\rho_c$ is the current critical density, and $\Omega_{\mathrm{HI}}$ is the comoving mass density in HI in units of the current critical density. This is assumed to take the fixed value $\Omega_{\mathrm{HI}} = 4.86 \times 10^{-4}$, consistent with its local value~\citep{2005MNRAS.359L..30Z}.

In Figure~\ref{fig:CHIME_SKA_EuclidELG_cov} we plot the contribution of interferometer noise to the dipole variance. For CHIME the Cosmic Variance $\times$ Interferometer Noise is the dominant source of variance for most pair separations, whereas for SKA it is subdominant due to the superior sensitivity. Similar conclusions apply to the Poisson $\times$ Interferometer Noise, which is only important for CHIME at low $d$. 

\subsection{Population weights}
\label{subsec:weights}

In situations where more than two populations of tracers may be distinguished, it will be possible to `stack' dipole signals from pairs of tracers to boost the signal-to-noise. Such an estimator would then take the form
\begin{equation}
\hat{\xi}(d) = \sum_{LM}h_{LM}(d) \hat{\xi}^{LM}(d),
\end{equation}
for some set of weights $h_{LM}$. The weights must obey the same symmetries as the correlation function estimator - for the dipole we must have antisymmetry in $(L,M)$. It is beyond the scope of this work to derive optimal weights for this purpose, but in~\citep{2016JCAP...08..021B} it was shown that in the case where Poisson noise is the dominant contribution to the multi-population covariance matrix, optimal weights are given by
\begin{equation}
h_{LM}(d) \propto \bar{n}_{L}\bar{n}_M \xi^{LM}(d).
\label{eq:hlm}
\end{equation}
Note that in the case where the slopes of the luminosity function vanish $s_A=s_B=0$, the weigths~\eqref{eq:hlm} simply reduce to the bias difference $\bar{n}_{L}\bar{n}_M (b_L- b_M)$. In most of the cases we consider in this work, Poisson noise is not the dominant source of variance (see Figure~\ref{fig:CHIME_SKA_EuclidELG_cov}), but nonetheless we shall see that adopting the weights of Equation~\eqref{eq:hlm} improves the signal-to-noise at each redshift over the naive $h_{LM} = [\Theta(F_L - F_M) - \Theta(F_M - F_L)]/2$ with $\Theta$ the Heaviside step function and $F$ some property of the tracer mapping onto the real numbers, such as luminosity. We refer to this choice of weighting as the `unweighted' estimator.

\section{Survey assumptions}
\label{sec:surveys}

In this section we describe the surveys included in our signal-to-noise forecasts. The numbers assumed here are only meant to be representative and are likely subject to flexibility, especially for the more advanced surveys.

\subsection{Number count surveys}

We will consider only galaxy surveys for which spectroscopic redshifts are available. In principle we could also consider photometric redshift surveys, which would require integrating signal and noise over broad redshift bins. Spectroscopic surveys suffer from having fewer tracers but their precise redshifts mean we can avoid integrating over a redshift distribution, which speeds up the calculation of the covariance matrix significantly. 

The parameters that need to be specified are then the number of redshift bins in the survey $N_z$ and their corresponding widths, the number of tracer populations $N_p$, the biases and magnification slopes in each bin, the number density $\bar{n}_g$ or total number $N_g$ of galaxies in each bin and the survey area $A_g$ or volume $V_g$.

\subsubsection{Euclid}

The Euclid mission~\citep{2011arXiv1110.3193L} will observe several tens of millions of near-infra red galaxies with spectroscopic redshifts obtained via the H$\alpha$ line. We follow Table 3 of~\citep{2016arXiv160600180A} and consider $N_z = 14$ redshift bins equally spaced between $z=0.65$ and $z=2.05$, assuming their `reference' case for the galaxy number densities. We will assume a survey area $A_g = 15,000\, \mathrm{sq. deg.}$.

Again following~\citep{2016arXiv160600180A}, the galaxy bias is taken as $b(z) = \sqrt{1+z}$. The magnification slope is more uncertain. In~\citep{2014JCAP...03..027C} an estimate of $s(m_*,z)$ was made using the projected H$\alpha$ counts of~\citep{2010MNRAS.402.1330G} for various different flux limits and redshift cuts. We adopt a value of $s(m_*,z) = 1.056$, consistent with that used in~\citep{2014JCAP...03..027C} assuming a flux limit of $1 \times 10^{-15.5} \, \mathrm{erg}\, \mathrm{cm}^{-2} \, \mathrm{s}^{-1}$. We assume the flux slope is constant with redshift, which is likely an incorrect assumption, but more detailed modelling is beyond the scope of this work - the redshift dependence is uncertain due to uncertainties in the redshift-evolution of the H$\alpha$ luminosity function. This is one of the reasons we do not consider the octupole in this work - it is sourced entirely by differences in the tracer magnification slopes, which are subject to large uncertainties. 

\subsubsection{DESI}

The Dark Energy Spectroscopic Instrument (DESI) will obtain spectra for roughly 18 million emission-line galaxies (ELGs), a few million luminous red galaxies (LRGs) and a few million quasi-stellar objects (QSOs). We take forecast parameters for the DESI survey from~\citep{2013arXiv1308.0847L}. We neglect QSOs due to their low number density, and instead consider ELGs and LRGs as distinct populations, i.e. $N_p = 2$. We also consider the case where all galaxies are used as a single population, i.e. $N_p = 1$.

We consider $N_z = 6$ redshift bins equally spaced between $z=0$ and $z=1.2$ for the $N_p=2$ case, and additionally include three higher-redshift bins out to $z=1.8$ in the $N_p = 1$ case, since these bins are expected to have ELGs but no LRGs (Table 3 of~\citep{2013arXiv1308.0847L}). We assume the survey covers an area $A_g = 14,000\, \mathrm{sq. deg.}$, and we take the bias of the tracers as $b(z) = b_0[D(z=0)/D(z)]$, where $D(z)$ is the growth factor and $b_0 = 1.7$ for LRGs and $b_0 = 0.84$ for ELGs. We fix the magnification slope as $s(m_*,z) = 1.0$ at every redshift for both ELGs and LRGs. As discussed above, there is considerable uncertainty as to what this value should be, and our choice should be considered a zeroth-order approximation. The number we have adopted is fairly close to the number calculated for Euclid ELGs in~\citep{2014JCAP...03..027C}.

\subsubsection{eBOSS}

The extended Baryon Oscillation Spectroscopic Survey (eBOSS) is a galaxy redshift survey currently underway as part of SDSS-IV. We take parameters from~\citep{2016MNRAS.457.2377Z}, and consider ELGs, LRGs, and QSOs as distinct tracers, i.e. $N_p = 3$. We consider $N_z = 4$ redshift bins equally spaced between $z=0.6$ and $z=1.0$, and additionally consider a scenario where all tracers are lumped together. In this case we include two extra bins out to $z=1.2$ for ELGs, and eight extra bins out to $z=2.2$ for QSOs.

Expected number densities are taken from Table 2 of~\citep{2016MNRAS.457.2377Z}, and the bias is taken as $b(z) = b_0[D(z=0)/D(z)]$ with $b_0 = 1.7$ for LRGs and $b_0 = 1.0$ for ELGs, and $b(z) = 0.53 + 0.29(1+z)^2$ for QSOs. We assume an area of $A_g = 1500\, \mathrm{sq. deg.}$. For the magnification slopes we assume $s(m_*,z) = 1.056$ for ELGs (i.e. the same as assumed for our Euclid ELG forecast), $s(m_*,z) = 1.2$ for LRGs (its value at $z=0.3$ for the BOSS survey according to~\citep{2009arXiv0901.1219M}), and $s(m_*,z) = 0.8$ for QSOs (as found for SDSS in~\citep{2005ApJ...633..589S}). Again we note that our assumptions regarding the magnification slopes are uncertain, but the numbers are broadly representative of previously measured values.

\subsubsection{BOSS}

As well as future and ongoing surveys, we also consider the completed Baryon Oscillation Spectroscopic Survey (BOSS), assuming $N_z = 2$ redshift bins corresponding to the LOWz and CMASS samples~\citep{2014MNRAS.441...24A}. We assume the sample may be split into two populations of bright and faint galaxies ($N_p = 2$), as has been done in~\citep{2015arXiv151203918G}, and use bias values derived for these populations in~\citep{2015arXiv151203918G}. Specifically we take $b=1.31$ for faint LOWz galaxies, $b=2.30$ for bright LOWz galaxies, $b=1.46$ for faint CMASS galaxies and $b=2.36$ for bright CMASS galaxies. We assume an area of $A_g = 4231\, \mathrm{sq. deg.}$ by averaging the survey areas of LOWz and CMASS. Finally, we fix $s(m_*,z) = 1.0$ for both bright and faint galaxies at each redshift.

\subsubsection{SKA}

Finally we consider Phase II of the Square Kilometre Array (SKA), a radio survey that will provide spectroscopic redshifts for tens to hundreds of millions of emission line galaxies over a large area of the sky. We take SKA specifications from~\citep{2015JCAP...10..070M, 2015aska.confE..21S}, and assume galaxies are binned into $N_z = 10$ redshift bins between $z=0.1$ and $z=2.0$, assuming the same bin distribution of~\citep{2015JCAP...10..070M}. The bins are all of roughly equal width except the highest redshift bin, which has $0.8 < z < 2.0$. For this high-redshift bin we approximate the signal and covariance matrix with their values at the mean redshift. The results for this bin are therefore only approximate - a proper treatment would involve integrating over the radial extend of the bin.

We assume the survey covers a fraction of sky $f_{\mathrm{sky}} = 0.73$, and the bias is taken as $b(z) = 0.5887\exp(0.813z)$, a fit from~\citep{2015aska.confE..21S} from which we also take the expected number densities. The magnification slopes are taken from the fitting formula of~\citep{2015MNRAS.448.1035C} assuming a flux limit of $5\, \mu \mathrm{Jy}$. Specifically we take $s(m_*,z) = 0.9329 - 1.5621\exp(-2.4377z)$. This is the only survey for which the magnification slope is assumed to vary with redshift, and should be taken as the most reliable forecast of this quantity.

\subsection{Intensity mapping surveys}

We consider two radio surveys with the capability of measuring 21\,cm brightness temperature fluctuations through intensity mapping. The parameters that need to be specified are the bias of 21\,cm emitters, their number density, and the noise parameters. Since we measure the total intensity in a frequency channel and the number density of sources depends on their intrinsic luminosity, our number density is really an \emph{effective} number density, given by an integral over flux weighted by the luminosity function of emitters~\citep[e.g.][]{1996MNRAS.281.1297T}. The magnification slope is fixed to $s(m_*,z) = 2/5$, as discussed in Section~\ref{sec:sig}. The only parameters that need to be specified further are the survey parameters entering the noise expressions of Section~\ref{sec:21noise}.

\subsubsection{CHIME}

The Canadian Hydrogen Intensity Mapping Experiment (CHIME) is a near-term survey designed to measure BAOs in the 21\,cm brightness temperature at high-redshifts ($0.8 < z < 2.5$). It consists of $N_{\mathrm{cyl}} = 5$ cylinders orientated in the North-South direction, each of length $L_{\mathrm{cyl}} = 100\mathrm{m}$ and width $W_{\mathrm{cyl}} = 20\mathrm{m}$, housing $N_f = 256$ dual-polarization feeds. We take parameters from~\citep{2014SPIE.9145E..4VN} and~\citep{2015ApJ...803...21B}, listed in Table~\ref{tab:21pars}. The effective area is calculated as $A_{\mathrm{eff}} = \eta L_{\mathrm{cyl}}W_{\mathrm{cyl}}N_{\mathrm{cyl}}/N_f$ where the efficiency $\eta = 0.7$.

We assume an isotropic FOV for each element. This is not correct in detail due to the cylindrical geometry of the array - elements are expected to have a FOV of roughly $90^{\circ}$ in the North-South direction and diffraction-limited in the East-West direction with an FOV of roughly $1.22\lambda/W_{\mathrm{cyl}}$. We take the product of the two directions to form our isotropic FOV.

\begin{table}
  \begin{tabular}{c c c}
      \hline
      \hline
       Parameter & \bf{CHIME} & \bf{SKA} \\
      & &  \\
      \hline
      $z_{\mathrm{min}}$ & 0.8 & 0.0 \\
      $z_{\mathrm{max}}$ & 2.5 & 3.06 \\
      $T_{\mathrm{ant}}$ & $50 \, \mathrm{K}$  & $20 \, \mathrm{K}$ ($z<0.58$) \\
       & &  $28 \, \mathrm{K}$ ($z>0.58$) \\
      $S_{21}$ & $25000 \, \mathrm{sq. deg.}$ &  $15000 \, \mathrm{sq. deg.}$     \\
      $N_f$ & 256 & 254\\
      $D_{\mathrm{max}}$ & 128m & 1000m\\
      $t_0$ & $10^4$ hrs & $10^4$ hrs \\
      $n_p$ & 2 & 2 \\
      \hline
      \end{tabular}
  \caption{Interferometer parameters used to compute the noise power spectrum of the 21\,cm noise. Parameters for CHIME are taken from~\citep{2014SPIE.9145E..4VN} and~\citep{2015ApJ...803...21B}, and those for SKA are taken from~\citep{2015ApJ...803...21B} and~\citep{2015aska.confE..21S}.}
  \label{tab:21pars}
  \end{table}
      
The bias of 21\,cm sources is assumed to take a constant value of $b_{21} = 0.6$. There is some uncertainty as to what this number should be and what redshift-dependence it should have, and the assumption of redshift-independence is common~\citep[e.g.][]{2010PhRvD..81f2001M} but almost certainly over-simplistic. Simulations suggest $b_{21}\approx 0.5 - 0.6$ around $z=1$, although this is subject to some model uncertainty~\citep{2011MNRAS.415.2580K}, and the bias could reach unity at these redshifts~\citep{2013ApJ...763L..20M}. If damped Lyman-$\alpha$ absorbers are responsible for the majority of 21\,cm emission at these redshifts, measurements from BOSS suggest a higher bias $b_{21} \approx 2$~\citep{2012JCAP...11..059F}. We take a compromise between these various predictions and fix $b_{21} = 0.6$. Note that the noise roughly scales with the bias, since the dominant contribution to the variance tends to involve the cosmic variance (see Figure~\ref{fig:CHIME_SKA_EuclidELG_cov}). The signal however depends on the bias \emph{difference} between the number counts and intensity map.

The effective number density of 21\,cm emitters is taken as $\bar{n}_{21} = 0.03 \, \mathrm{Mpc}^{-3}\,h^{3}$, assumed constant with redshift, following~\citep{2010PhRvD..81f2001M}.

\subsubsection{SKA}

Whilst CHIME offers the potential for the near-term intensity mapping across a large area of the sky, the redshifts probed are too high for cross-correlation studies with many of the bins for which number counts are available. We therefore additionally consider forecast intensity maps from the SKA, which in principle will observe most of the sky down to $z\approx 0$.

Specifically, we will assume parameters from~\citep{2015ApJ...803...21B} corresponding to SKA Phase I in `interferometer' mode. We assume a total of $N_f = 254$ feeds, 190 of which come from the SKA MID array and 64 from the MeerKAT array. The effective area of the elements is taken as $A_{\mathrm{eff}} = 140\,\mathrm{m}^2$~\citep{2015aska.confE..19S}, and the sky area as $S_{21} = 15000\,\mathrm{sq. deg.}$. We combine the two frequency bands B1 and B2, such that the survey probes redshifts $0.00 < z < 3.06$, and the antenna temperature is taken as $T_{\mathrm{ant}} = 20\,\mathrm{K}$ for the higher frequency B2 band and $T_{\mathrm{ant}} = 28\,\mathrm{K}$ for the lower frequency B1 band. Finally we assume a maximum baseline of $D_{\mathrm{max}} = 1000\,\mathrm{m}$, dual-polarization feeds ($n_p = 2$), and the same total observing time as CHIME, $t_0 = 10^4 \, \mathrm{hrs}$. These parameters are listed in the second column of Table~\ref{tab:21pars}. We assume the number density and bias of 21\,cm emitters are the same as for the CHIME experiment.

Since the SKA can be used as both a galaxy number counts survey and a 21\,cm intensity mapping survey, the possibility exists that the discrete tracers observed in the number counts survey make up the bulk of the total 21\,cm intensity. If this is the case, the biases of the two sources should be the same, but the magnification slopes will necessarily differ due to the different observables involved and hence the cross-correlation dipole will be non-zero. We refer to this situation as the `common tracer' case, and compute separate forecasts for this scenario. The Poisson noise should also be modified in this case, since the tracers responsible are identified~\citep{2009MNRAS.400..851S}. The bias and number density of 21\,cm emitters in the common tracer case are assumed to be the same as for the number counts survey, i.e. evolving with redshift.

\section{Forecasts}
\label{sec:forecasts}

Now that we have defined our estimators, noise and survey parameters, we can ask the question of how detectable the cross-correlation dipole will be with current and future experiments. To answer this, we first define the signal-to-noise ($S/N$) of the measurement as
\begin{equation}
(S/N)(z) = \sqrt{\sum_{d,d'} \xi_1(d;z) C^{-1}(d,d';z) \xi_1(d';z)},
\end{equation}
where the expected cross-correlation dipole and its covariance matrix have been summed over populations with the weights of Section~\ref{subsec:weights}. The minimum $d$ in the above summation is chosen as either four times the angular resolution of the 21\,cm survey or the non-linear scale $d_{\mathrm{nl}}$, whichever is greater. We use the expression~\citep{2015ApJ...803...21B}
\begin{equation}
d_{\mathrm{nl}} = 5(1+z)^{[-2/(2+n_s)]} \, \mathrm{Mpc}\,h^{-1}.
\end{equation}
The maximum value of $d$ is taken to be the maximal distance available at all orientations in the overlap region between the surveys, or the quantity $r/10$, whichever is smaller. This latter condition is to ensure that wide-angle terms are kept subdominant in both the signal and covariance. We bin the correlation function in $d$ with bin-width given by the angular resolution of the 21\,cm survey.

\subsection{Single population galaxy surveys}

All our $S/N$ forecasts are tabulated in Appendix~\ref{app:sn}. In Figure~\ref{fig:CHIME_SN_sp} we plot the forecast $S/N$ on the dipole for correlations between CHIME and single-population number counts. For DESI we lump ELGs and LRGs together, while for eBOSS we consider the separate populations individually. CHIME has relatively high interferometer noise, but the dipole is still detectable at the 2$\sigma$-level for the lowest redshift bin when combined with DESI. A positive detection at low redshift may also be made with Euclid, albeit at the lower significance of roughly $1.5\sigma$. More promising is the correlation of CHIME with SKA number counts - in this case a $2.5\sigma$ detection is possible, although we caution the reader that this result computes the $S/N$ at the mean redshift of the broad high-redshift bin of the SKA survey, which is only a very approximate treatment as discussed above. The eBOSS survey does not have a large enough area to beat down Cosmic Variance sufficiently to make a measurement of the dipole, and the current BOSS survey is at too low a redshift to be combined with CHIME.

The sharp decrease in $S/N$ with redshift is due to a combination of the signal decreasing due to the amplitude of the power spectrum decreasing (which overcomes the fact that the bias-difference tends to increase with redshift, see Figure~\ref{fig:CHIME_EuclidELG_SIG}), the lower $d_{\mathrm{max}}$ available due the narrower conformal width of the higher redshift bins, the higher $d_{\mathrm{min}}$ at higher redshift due to the fixed angular resolution, and to the weak dependence of the dominant Cosmic Variance $\times$ Interferometer Noise on redshift. This latter feature is due to the decrease in the cosmic variance part of the noise with redshift being largely compensated by the increase in the interferometer noise with redshift. The case of eBOSS QSOs is somewhat different - the dominant noise term here is the Poisson $\times$ Interferometer Noise term, and the bias rises steeply with redshift. These effects combine to give a flatter trend of $S/N$ with redshift.

\begin{figure}
\centering
\includegraphics[width=0.6\textwidth]{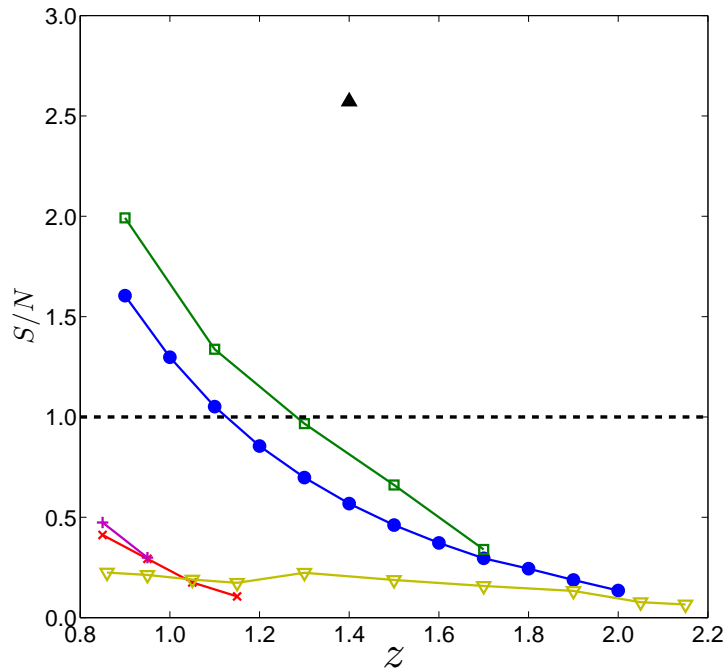}
\caption{(Colour Online). Forecast $S/N$ values for correlations between CHIME and single-population galaxy surveys. Shown are values for correlations between CHIME and Euclid ELGs (blue circles), DESI combined ELGs and LRGs (green squares), eBOSS ELGs (red crosses), eBOSS LRGs (magenta plus signs), eBOSS QSOs (yellow down-pointing triangles), and SKA (black upward-pointing triangle). The horizontal black dashed line indicates a $S/N$ of unity.}
\label{fig:CHIME_SN_sp}
\end{figure}

\begin{figure}
\centering
\includegraphics[width=0.6\textwidth]{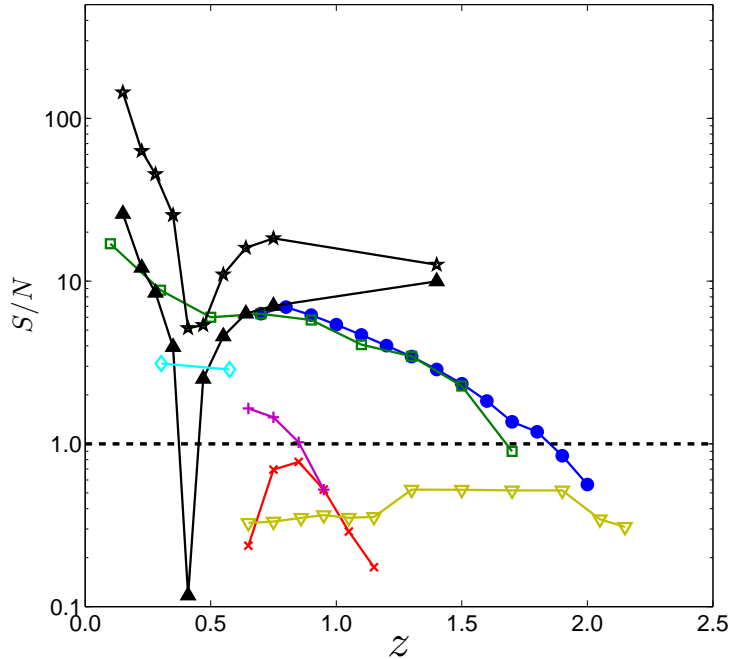}
\caption{(Colour Online). Forecast $S/N$ values for correlations between SKA and single-population galaxy surveys. Shown are values for correlations between SKA and Euclid ELGs (blue circles), DESI combined ELGs and LRGs (green squares), eBOSS ELGs (red crosses), eBOSS LRGs (magenta plus signs), eBOSS QSOs (yellow down-pointing triangles), BOSS (cyan diamonds), SKA assuming distinct tracers (black upward-pointing triangles), and SKA assuming common tracers (black stars). The horizontal black dashed line indicates a $S/N$ of unity.}
\label{fig:SKA_SN_sp}
\end{figure}

In Figure~\ref{fig:SKA_SN_sp} we plot forecasts for SKA intensity mapping with single-population number counts surveys. The high sensitivity of the SKA combined with its high resolution and huge survey volume mean that a significant detection of the cross-correlation dipole looks likely. A $S/N > 5$ is achieved at low redshift for both Euclid and DESI. 

Very high $S/N$ values are achievable using correlations completely internal to the SKA, i.e. correlating SKA number counts with intensity mapping. This is due to both an enhancement of the signal and, more importantly, a considerable suppression of the noise. The former arises due to the fact that the first and second terms in braces in the second line of Equation~\eqref{eq:pmults} tend to cancel each other, and when $b_A = b_B$ this results in a roughly $50\%$ increase in the signal on all scales. The lower noise is due to a significant suppression of both the Cosmic Variance $\times$ Poisson and Poisson $\times$ Poisson contributions to the noise, which are the dominant terms for the SKA. This can be seen by inspection of Equation~\eqref{eq:covmat}. For dipolar weights and common tracers, both Poisson terms in the last line are non-zero, and exactly cancel. Similarly, each of the four terms in the second and third lines are non-zero, with their sum proportional to $P_L^{BB} + P_L^{AA} - 2P_L^{BA}$ with $L=0,2$. When the biases are equal these terms cancel. Hence there is a significant increase in the forecast $S/N$ when common tracers are assumed for the SKA.

The interesting exception to this is around $z\approx 0.4$, where the $S/N$ forecast dips sharply. The reason for this is that the cross-correlation dipole undergoes a sign-change around this redshift. This can be traced to the part of the signal coming from the magnification slope - around $z\approx 0.4$ the quantity $s(m_*,z) - s_{21} = 0.9329 - 1.5621\exp(-2.4377z) - 2/5$ changes sign. In the case where the tracers are assumed to be common (upper black solid line and stars in Figure~\ref{fig:SKA_SN_sp}), the entirety of the signal comes from this term, and the $S/N$ goes exactly to zero (not visible on this plot due to the binning). This serves to illustrate an important cautionary note regarding the forecasts at low redshift. At low redshift the term in Equation~\eqref{eq:numcount_euler} proportional to $s(m_*,z)/(\mathcal{H}\chi)$ (familiar from~\citep{2008MNRAS.389..292P}) can become large, making our prediction for the dipole quite sensitive to the poorly forecast value of $s(m_*,z)$. This is evident in the strong dependence of the low-redshift SKA forecasts on the sign-change of $s(m_*,z) - s_{21}$, and suggests these forecasts should be treated with some caution. In Section~\ref{subsec:systematics} we quantify the sensitivity of our forecasts to the magnification slope.

The trends of the $S/N$ forecasts with redshift are also somewhat non-trivial for the other number-counts surveys considered. Figure~\ref{fig:CHIME_SKA_EuclidELG_cov} suggests that the Cosmic Variance $\times$ Poisson noise is the dominant noise term, at least for Euclid ELGs. This drops with redshift due to the decrease in the amplitude of the matter power spectrum, which overcomes the fact that the physical pixel scale $L_p$ increases with redshift at fixed angular resolution. 

\subsection{Multiple population galaxy surveys}

\begin{figure}
\centering
\includegraphics[width=0.6\textwidth]{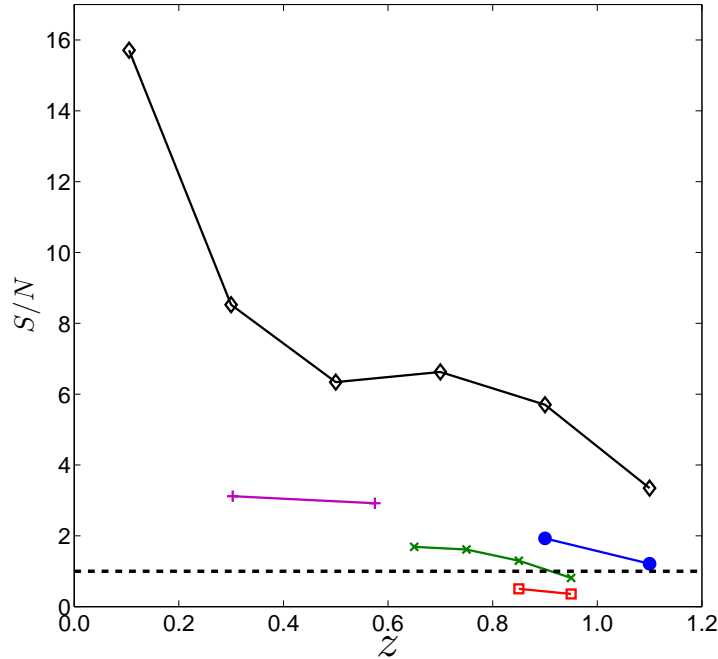}
\caption{(Colour Online). Forecast $S/N$ values for correlations between SKA and CHIME with multiple-population galaxy surveys, assuming the weights of Equation~\eqref{eq:hlm} are applied. Shown are values for correlations between CHIME and DESI (blue circles), CHIME and eBOSS (red squares), SKA and DESI (black diamonds), SKA and eBOSS (green crosses), and SKA and BOSS (magenta plus signs). The horizontal black dashed line indicates a $S/N$ of unity.}
\label{fig:SN_mp}
\end{figure}

In Figure~\ref{fig:SN_mp} we plot forecast $S/N$ values for correlations between both CHIME and SKA intensity mapping with the three surveys we consider having multiple galaxy populations - BOSS, eBOSS, and DESI. We assume that the Poisson-optimised weights of Equation~\eqref{eq:hlm} have been applied. The results are very encouraging, with a $S/N$ above unity achievable at most redshifts and for most combinations of surveys. The most significant detections of the dipole come from SKA combined with DESI, and detections are also possible with combinations of SKA with BOSS and eBOSS, and with CHIME combined with DESI. A detection with CHIME combined with eBOSS does not look so promising.

As quantified in Table~\ref{table:mpsn} of Appendix~\ref{app:sn}, stacking the signal from multiple pairs of tracers with the Poisson-optimised weights of Equation~\eqref{eq:hlm} always improves the $S/N$ over the naive weights discussed in Section~\ref{subsec:weights}. Depending on how important Poisson noise is in the covariance matrix, the improvement can either be modest or significant. For example, in the highest redshift bin of the SKA $\times$ DESI cross-correlation there are expected to be very few LRGs, and hence Poisson-optimised weights can boost the $S/N$ from 0.79 to 3.35. Typical improvements range between tens of percent to factors of a few. 

In most cases, the Poisson-optimised weights are required to ensure stacking the populations does not \emph{degrade} the $S/N$ over the individual tracer combinations - the naive weights correspond to throwing away information by averaging the signal. In most cases, use of the weights of Equation~\eqref{eq:hlm} gives a higher $S/N$ than the highest $S/N$ given by combining the individual populations (Tables~\ref{table:chimesn} and~\ref{table:skasn}). The only exceptions to this are the central bins of the SKA $\times$ DESI cross-correlation, where lumping the ELGs and LRGs into one population gives a higher $S/N$ than stacking the individual populations with the Poisson-optimised weights. Although the degradation is mild, this demonstrates that in situations where the variance is not dominated by Poisson noise, the weights of Equation~\eqref{eq:hlm} may be far from optimal.

\begin{figure}
\centering
\includegraphics[width=0.6\textwidth]{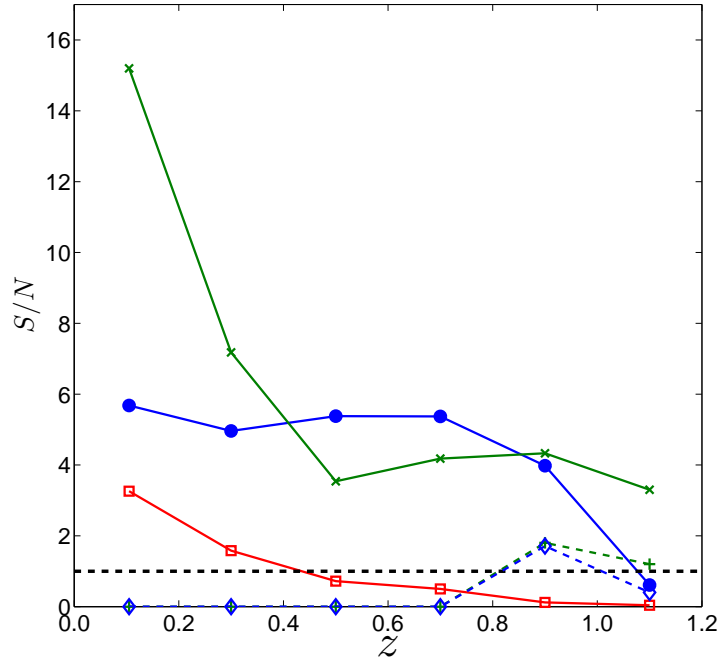}
\caption{(Colour Online). Pairwise $S/N$ for DESI with CHIME and SKA. Shown are values for correlations between DESI ELGs and LRGs (red squares), CHIME and DESI ELGs (green dashed, plus signs), SKA and DESI ELGs (green solid, crosses), CHIME and DESI LRGs (blue dashed, diamonds), SKA and DESI LRGs (blue solid, circles). The horizontal black dashed line indicates a $S/N$ of unity.}
\label{fig:DESI_pwise}
\end{figure}

We can also ask the question of what the inclusion of 21\,cm intensity maps brings to the measurement of the dipole. For surveys with multiple populations the signal could be estimated from the individual galaxy populations without including 21\,cm, and given that interferometer noise can dominate the estimator variance it might be worried that including intensity maps brings little to the table, especially for galaxy surveys having spectroscopic redshifts available. In Figure~\ref{fig:DESI_pwise} we plot the forecast $S/N$ for DESI combined with SKA and CHIME, showing the various combinations of ELGs, LRGs, and 21\,cm intensity mapping. This plot clearly demonstrates that at all redshifts where 21\,cm intensity maps can be measured they boost the signal when combined with ELGs or LRGs (blue and green curves) over what can be achieved by combining ELGs with LRGs (red curve). This is due to a combination of factors such as the tendency for intensity maps to cover larger areas of the sky, the more dramatic difference between the magnification slopes when comparing galaxies and the brightness temperature, and the higher number density of 21\,cm emitters. We would expect the benefits of including 21\,cm intensity maps to be greater when combined with number counts having poorly constrained redshifts or small survey volumes.

To conclude this section, we have seen that the cross-correlation dipole should be measurable at modest significance with CHIME when combined with DESI or Euclid at low redshift, and with high significance when combining the SKA with number counts from Euclid, DESI, or itself. The signal does not appear to be measurable with either eBOSS or BOSS, even when combined with SKA.

\subsection{Non-linear effects and sensitivity to the flux slopes.}
\label{subsec:systematics}

\begin{figure}
\centering
\includegraphics[width=0.6\textwidth]{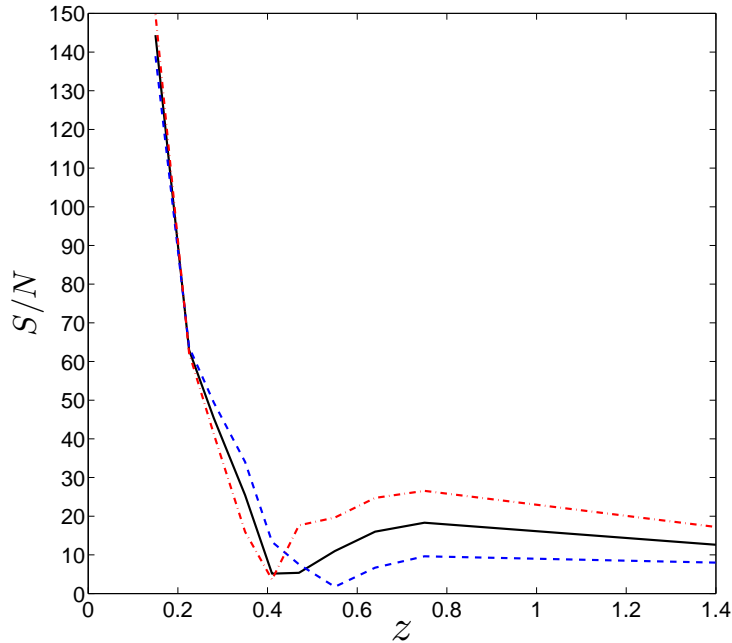}
\caption{(Colour Online). Effect of varying the flux counts slope on the forecast $S/N$ for the SKA 21\,cm and SKA galaxy survey cross-correlation dipole, assuming common tracers, as a function of redshift. We vary the flux slope $s(m_*,z)$ by $\pm 20\%$ at each redshift (red dot-dashed and blue dashed respectively).}
\label{fig:SKA_SKA_skolp_varmag}
\end{figure}

Finally, there are a few simplifying assumptions in this work that merit some discussion. Firstly, we have seen that some of our forecasts are sensitive to the poorly forecast values of the magnification slope $s(m_*,z)$. To get an idea of how sensitive our results are to this quantity, in Figure~\ref{fig:SKA_SKA_skolp_varmag} we plot the forecast $S/N$ for SKA 21\,cm intensity maps combined with SKA number counts, assuming common tracers, and varying $s(m_*,z)$ by $\pm 20\%$. This changes the redshift at which the signal changes sign and changes the $S/N$ at higher redshift by roughly $\pm 50\%$. At the lowest redshifts where the $S/N$ is highest however, the effect is only small, with the $S/N$ being determined more by the bias. 

The dramatic rise in $S/N$ in Figure~\ref{fig:SKA_SKA_skolp_varmag} as the redshift gets small is due primarily to the smaller $d_{\mathrm{min}}$ allowed by the fixed angular resolution. We thus might worry that these lower bins might be sensitive to enhancements in the signal coming from non-linear clustering in the matter power spectrum. We have partly accounted for this by ensuring the lower $d$ used is never smaller than a characteristic non-linear scale $d_{\mathrm{nl}}$ (see Section~\ref{sec:forecasts}), but this only approximately accounts for non-linearity, since the $\ell=1$ Bessel functions in our cross-correlation integrals fall off slowly with $kd$, meaning that non-linear power could contribute to pair separations greater than $d_{\mathrm{nl}}$. To test the effects of this, we re-ran all our $S/N$ forecasts with a modified power spectrum including non-linear power using the \textsc{HALOFIT} package~\citep{2012ApJ...761..152T}~\footnote{Note that this procedure gives only an approximate description of non-linear effects, since it relies on using linear Einstein's equations to relate the velocity to the density. A more accurate procedure would be to model the non-linear velocity directly, using for example the streaming model, see e.g.~\cite{2013MNRAS.431.2834X}. However, since our procedure tends to over-estimate the effect of non-linearities on the velocity, we can regard it as a conservative upper bound.}. We find that our results are robust to including non-linear power, with forecast $S/N$ values increasing only by a few percent in most cases. The insensitivity to small scales is due to our resolution-scale smoothing and choice of $d$-cuts, the largest effects being at low redshift where the minimum $d$ tends to be lower and non-linear enhancement to the matter power spectrum is greater.

\section{Conclusions}
\label{sec:concs}
In this work we have investigated the feasibility of measuring a relatively unexplored observable of large-scale structure, the dipole of the multi-tracer redshift-space correlation function. This quantity is sensitive to contributions to the observed number density that cannot be probed with the standard even multipoles, and provides complementary information on large-scale bulk motion to that provided by redshift-space distortions. This information is however contaminated by wide-angle effects, which dominate current measurements of the dipole~\citep{2015arXiv151203918G}, and must be removed to isolate the desired Doppler terms.

We have derived a new expression for the covariance matrix of multi-tracer redshift-space correlation functions in the Gaussian limit, which should be useful for future surveys aiming at conducting a multi-tracer analysis such as eBOSS or DESI. In an upcoming work we will test the robustness of our expression on available data from BOSS~\citep{inprep}. We also account for the extra variance incurred by removing the wide-angle term, finding it to be negligible in most cases due to a partial cancellation of cross and auto terms.

Using measured and expected values for biases, magnification slopes, number densities and other survey parameters, we have forecast $S/N$ values for the amplitude of the dipole as a function of redshift. The most significant detection is likely to come from the SKA using its number counts survey combined with its 21\,cm intensity mapping survey, with $S/N \lesssim 100$ possible at low redshift. The significance is enhanced when these two tracers are identified due to a significant suppression of the Poisson noise, although there is a redshift ($z\approx0.4$) where the $S/N$ drops to zero for the SKA, due to the sign of the signal changing.

Detections are also possible with SKA combined with the future galaxy surveys DESI and Euclid ($S/N \lesssim 10$), and potentially with the near-term radio survey CHIME when combined with DESI and Euclid ($S/N \lesssim 2$).

The significance of the dipole may be enhanced by stacking multiple populations. We have not attempted to derive optimal weights, but even for Poisson-optimal weights the significance may be boosted over the single-population case. For the SKA we expect the dipole to be measured with $S/N \approx 10$ when combined with DESI.

We have tested the robustness of our results to the uncertain magnification slope, finding a strong dependence at moderate redshift to high redshift ($z \gtrsim 0.5$), and a fairly weak dependence at lower redshifts. Our results are also insensitive to non-linear corrections to the power spectrum, although we have not included non-linear corrections to the covariance matrix, which would enhance our error bars.

Finally, we have seen that the inclusion of 21\,cm intensity maps genuinely adds something to the measurement of the dipole. The large volume accessible with radio surveys combined with the expected high number density of tracers and fundamentally different magnification slopes imply that combining the 21\,cm brightness temperature with number counts from galaxy surveys can be an excellent way to measure the dipole. Potential caveats to this include the removal of the unknown sky-average brightness temperature, which will require the inclusion of the 21\,cm auto-correlation. This will add some variance to our estimator which will need to be accounted for. We have also neglected foregrounds in this work - although these do not bias the estimator they will contribute to the variance. We have assumed that intensity maps have been cleaned of foregrounds, which should be possible in principle due to their distinct frequency dependence~\citep[e.g.][]{2009MNRAS.398..401L}.

In conclusion, a detection of the dipole as a function of redshift is seemingly within reach in the next decade. Such a measurement will provide a very interesting consistency check on the $\Lambda$CDM model and will also provide complementary constraints on the equivalence principle and scale-dependent growth, allowing for tests of novel extensions to the standard model, such as a modified theory of gravity. Information from 21\,cm intensity mapping surveys will be invaluable in this endeavour.

\section{Acknowledgments}
AH is supported by a United Kingdom Space Agency \emph{Euclid} grant, and an STFC Consolidated Grant. AH thanks the CERN Theory Division for their hospitality and acknowledges a visitors grant. CB acknowledges support by the Swiss National Science Foundation.


\appendix

\section{Tabulated $S/N$ forecasts}
\label{app:sn}

\begin{table}
  \begin{tabular*}{0.75\textwidth}{@{\extracolsep{\fill}}c c c c}
      \hline
      \hline
      \bf{Radio Survey} & \bf{Galaxy Survey} & $z$\bf{-bin} & $S/N$ \\
      \hline
      CHIME & Euclid (ELGs)  & $0.85 < z < 0.95 $ & 1.60 \\
            &             & $0.95 < z < 1.05 $ & 1.30 \\
            &             & $1.05 < z < 1.15 $ & 1.05 \\
            &             & $1.15 < z < 1.25 $ & 0.86 \\
            &             & $1.25 < z < 1.35 $ & 0.70 \\
            &             & $1.35 < z < 1.45 $ & 0.57 \\
            &             & $1.45 < z < 1.55 $ & 0.46 \\
            &             & $1.55 < z < 1.65 $ & 0.37 \\
            &             & $1.65 < z < 1.75 $ & 0.30 \\
            &             & $1.75 < z < 1.85 $ & 0.24 \\
            &             & $1.85 < z < 1.95 $ & 0.19 \\
            &             & $1.95 < z < 2.05 $ & 0.13 \\
      \hline
      CHIME & DESI (ELGs + LRGs) & $0.8 < z < 1.0$ & 2.00 \\
            &      & $1.0 < z < 1.2$ & 1.34 \\
            &      & $1.2 < z < 1.4$ & 0.97 \\
            &      & $1.4 < z < 1.6$ & 0.66 \\
            &      & $1.6 < z < 1.8$ & 0.34 \\
      \hline
      CHIME & eBOSS (ELGs)  & $0.8 < z < 0.9$ & 0.41 \\
            &       & $0.9 < z < 1.0$ & 0.29 \\
            &       & $1.0 < z < 1.1$ & 0.18 \\
            &       & $1.1 < z < 1.2$ & 0.11 \\
      \hline
      CHIME & eBOSS (LRGs) & $0.8 < z < 0.9$ & 0.47 \\
            &       & $0.9 < z < 1.0$ & 0.30 \\
      \hline
      CHIME & eBOSS (QSOs) & $0.8 < z < 0.9$ & 0.22 \\
            &       & $0.9 < z < 1.0$ & 0.21 \\
            &       & $1.0 < z < 1.1$ & 0.19 \\
            &       & $1.1 < z < 1.2$ & 0.17 \\
            &       & $1.2 < z < 1.4$ & 0.22 \\
            &       & $1.4 < z < 1.6$ & 0.19 \\
            &       & $1.6 < z < 1.8$ & 0.16 \\
      \hline
      CHIME & SKA & $0.81 < z < 1.99$ & 2.57 \\
      \hline
      \hline
  \end{tabular*}
  \caption{$S/N$ forecasts for CHIME and individual galaxy populations.}
  \label{table:chimesn}
\end{table}

\begin{table}
  \begin{tabular*}{0.75\textwidth}{@{\extracolsep{\fill}}c c c c}
      \hline
      \hline
      \bf{Radio Survey} & \bf{Galaxy Survey} & $z$\bf{-bin} & $S/N$ \\
      \hline
      SKA & Euclid (ELGs) & $0.65 < z < 0.75 $ & 6.32 \\
            &             & $0.75 < z < 0.85 $ & 6.93 \\
            &             & $0.85 < z < 0.95 $ & 6.18 \\
            &             & $0.95 < z < 1.05 $ & 5.40 \\
            &             & $1.05 < z < 1.15 $ & 4.67 \\
            &             & $1.15 < z < 1.25 $ & 4.02 \\
            &             & $1.25 < z < 1.35 $ & 3.44 \\
            &             & $1.35 < z < 1.45 $ & 2.86 \\
            &             & $1.45 < z < 1.55 $ & 2.34 \\
            &             & $1.55 < z < 1.65 $ & 1.83 \\
            &             & $1.65 < z < 1.75 $ & 1.37 \\
            &             & $1.75 < z < 1.85 $ & 1.19 \\
            &             & $1.85 < z < 1.95 $ & 0.85 \\
            &             & $1.95 < z < 2.05 $ & 0.56 \\
      \hline
      SKA & DESI (ELGs + LRGs) & $0.0 < z < 0.2$ & 16.7 \\
            &      & $0.2 < z < 0.4$ & 8.80 \\
            &      & $0.4 < z < 0.6$ & 6.00 \\
            &      & $0.6 < z < 0.8$ & 6.29 \\
            &      & $0.8 < z < 1.0$ & 5.77 \\
            &      & $1.0 < z < 1.2$ & 4.08 \\
            &      & $1.2 < z < 1.4$ & 3.45 \\
            &      & $1.4 < z < 1.6$ & 2.23 \\
            &      & $1.6 < z < 1.8$ & 0.90 \\
      \hline
      SKA & eBOSS (ELGs) & $0.6 < z < 0.7$ & 0.25 \\
            &       & $0.7 < z < 0.8$ & 0.69 \\
            &       & $0.8 < z < 0.9$ & 0.77 \\
            &       & $0.9 < z < 1.0$ & 0.52 \\
            &       & $1.0 < z < 1.1$ & 0.29 \\
            &       & $1.1 < z < 1.2$ & 0.17 \\
      \hline
      SKA & eBOSS (LRGs) & $0.6 < z < 0.7$ & 1.65 \\
            &       & $0.7 < z < 0.8$ & 1.46 \\
            &       & $0.8 < z < 0.9$ & 1.02 \\
            &       & $0.9 < z < 1.0$ & 0.52 \\
      \hline
      SKA & eBOSS (QSOs) & $0.6 < z < 0.7$ & 0.33 \\
            &       & $0.7 < z < 0.8$ & 0.33 \\
            &       & $0.8 < z < 0.9$ & 0.35 \\
            &       & $0.9 < z < 1.0$ & 0.36 \\
            &       & $1.0 < z < 1.1$ & 0.35 \\
            &       & $1.1 < z < 1.2$ & 0.36 \\
            &       & $1.2 < z < 1.4$ & 0.52 \\
            &       & $1.4 < z < 1.6$ & 0.52 \\
            &       & $1.6 < z < 1.8$ & 0.52 \\
      \hline
      SKA & BOSS & $0.15 < z < 0.42$ & 3.12 \\
          &      & $0.44 < z < 0.70$ & 2.86 \\
      \hline
      SKA & SKA  & $0.1 < z < 0.2$ & 25.9\, / \,144.4 \\
            &     & $0.2 < z < 0.25$ & 12.1\, / \,63.0 \\
            &     & $0.25 < z < 0.31$ & 8.49\, / \,45.4 \\
            &     & $0.31 < z < 0.39$ & 3.94\, / \,25.4 \\
            &     & $0.39 < z < 0.43$ & 0.12\, / \,5.14 \\
            &     & $0.43 < z < 0.51$ & 2.51\, / \,5.37 \\
            &     & $0.51 < z < 0.59$ & 4.58\, / \,11.0 \\
            &     & $0.59 < z < 0.69$ & 6.31\, / \,16.0 \\
            &     & $0.69 < z < 0.81$ & 7.04\, / \,18.3 \\
            &     & $0.81 < z < 1.99$ & 9.96\, / \,12.6 \\
      \hline
      \hline
      \end{tabular*}
  \caption{$S/N$ forecasts for SKA and individual galaxy populations. In the case of SKA 21\,cm combined with SKA number counts we show the $S/N$ assuming distinct and common tracers respectively.}
  \label{table:skasn}
  \end{table}

\begin{table}
  \begin{tabular*}{0.75\textwidth}{@{\extracolsep{\fill}}c c c c c}
      \hline
      \hline
      \bf{Radio Survey} & \bf{Galaxy Survey} & $z$\bf{-bin} & $S/N$ \bf{(weighted)} & $S/N$\bf{(unweighted)} \\
      \hline
      CHIME & DESI & $0.8 < z < 1.0$ & 1.93 & 1.70 \\
            &      & $1.0 < z < 1.2$ & 1.21 & 0.54 \\
      \hline
      CHIME & eBOSS & $0.8 < z < 0.9$ & 0.50 & 0.33 \\
            &       & $0.9 < z < 1.0$ & 0.36 & 0.24 \\
      \hline
      SKA & DESI & $0.0 < z < 0.2$ & 15.7 & 12.9 \\
          &      & $0.2 < z < 0.4$ & 8.52 & 5.85 \\
          &      & $0.4 < z < 0.6$ & 6.34 & 2.82 \\
          &      & $0.6 < z < 0.8$ & 6.63 & 3.16 \\
          &      & $0.8 < z < 1.0$ & 5.70 & 2.93 \\
          &      & $1.0 < z < 1.2$ & 3.35 & 0.79 \\
      \hline
      SKA & eBOSS & $0.6 < z < 0.7$ & 1.69 & 0.17 \\
          &       & $0.7 < z < 0.8$ & 1.61 & 0.38 \\
          &       & $0.8 < z < 0.9$ & 1.30 & 0.39 \\
          &       & $0.9 < z < 1.0$ & 0.81 & 0.27 \\
      \hline
      SKA & BOSS & $0.15 < z < 0.42$ & 3.12 & 1.49 \\
          &      & $0.44 < z < 0.70$ & 2.92 & 1.16 \\
      \hline
      \hline
      \end{tabular*}
  \caption{$S/N$ forecasts for galaxy surveys having multiple populations.}
  \label{table:mpsn}
  \end{table}

\end{document}